\documentclass[usenatbib]{mn2e}
\usepackage{graphicx}
\usepackage{ifthen}
\usepackage{url}
\usepackage{amsmath}
\usepackage{rotating}

\def\ltsima{$\; \buildrel < \over \sim \;$}
\def\lta{\lower.5ex\hbox{\ltsima}}
\def\gtsima{$\; \buildrel > \over \sim \;$}
\def\simgt{\lower.5ex\hbox{\gtsima}}
\def\kms{{\rm\,km \; s^{-1}}}

\def\masyr{{\rm\,mas/yr}}
\def\kpc{{\rm\,kpc}}

\def\msun{{\rm\,M_\odot}}
\def\lsun{{\rm\,L_\odot}}

\def\pc{{\rm\,pc}}

\def\AA{$\; \buildrel \circ \over {\rm A}$}

\def\s{\ifmmode \widetilde \else \~\fi}
\def\={\overline}

\def\spose#1{\hbox to 0pt{#1\hss}}

\def\eg{{e.g.,\ }}

\def\lta{\mathrel{\spose{\lower 3pt\hbox{$\mathchar"218$}}
     \raise 2.0pt\hbox{$\mathchar"13C$}}}
\def\gta{\mathrel{\spose{\lower 3pt\hbox{$\mathchar"218$}}
     \raise 2.0pt\hbox{$\mathchar"13E$}}}
\def\Dt{\spose{\raise 1.5ex\hbox{\hskip3pt$\mathchar"201$}}}    
\def\dt{\spose{\raise 1.0ex\hbox{\hskip2pt$\mathchar"201$}}}    

\def\dotsfill{\leaders\hbox to 1em{\hss.\hss}\hfill}

\def\Gyr{{\rm\,Gyr}}
\def\FeH{{\rm[Fe/H]}}

\loadboldmathitalic 
\title[Pristine Dwarf-Galaxy Survey I]{Pristine Dwarf-Galaxy Survey I: A detailed photometric and spectroscopic study of the very metal-poor Draco II satellite.}
\author[N. Longeard et al.] {Nicolas Longeard$^{1}$, Nicolas Martin$^{1,2}$, Else Starkenburg$^{3}$, Rodrigo A. Ibata$^{1}$, 
\newauthor Michelle L. M. Collins$^{4,6}$, Marla Geha$^{6}$, Benjamin P. M. Laevens$^{5}$, R. Michael Rich$^{7}$, 
\newauthor David S. Aguado$^{8,9}$, Anke Arentsen$^{3}$, Raymond G. Carlberg$^{10}$, Patrick C\^ot\'e$^{11}$,
\newauthor Vanessa Hill$^{12}$, Pascale Jablonka$^{13,14}$,  Jonay I. Gonz\'alez Hern\'andez$^{8,9}$, 
\newauthor Julio F. Navarro$^{15}$, Rub\'en S\'anchez-Janssen$^{16}$, Eline Tolstoy$^{17}$,  Kim A. Venn$^{15}$,
\newauthor Kris Youakim$^{3}$ \\
$^{1}$ Universit\'e de Strasbourg, CNRS, Observatoire astronomique de Strasbourg, UMR 7550, F-67000 Strasbourg, France\\
$^{2}$ Max-Planck-Institut f\"ur Astronomy, K\"onigstuhl 17, D-69117, Heidelberg, Germany\\
$^{3}$ Leibniz Institute for Astrophysics Potsdam (AIP), An der Sternwarte 16, 14482 Potsdam, Germany\\
$^{4}$ Department of Physics, University of Surrey, Guildford, GU2 7XH, Surrey, UK\\
$^{5}$ Institute of Astrophysics, Pontificia Universidad Cat\'olica de Chile, Av. Vicuña Mackenna 4860, 7820436 Macul, Santiago, Chile\\
$^{6}$ Department of Astronomy, Yale University, New Haven, CT 06520, USA\\
$^{7}$ University of California Los Angeles, Department of Physics \& Astronomy, Los Angeles, CA, USA\\
$^{8}$ Instituto de Astrofisica de Canarias, Via Lactea, 38205 La Laguna, Tenerife, Spain\\
$^{9}$ Universidad de La Laguna, Departamento de Astrofisica, 38206 La Laguna, Tenerife, Spain\\
$^{10}$ Department of Astronomy \& Astrophysics, University of Toronto, Toronto, ON M5S 3H4, Canada\\
$^{11}$ NRC Herzberg Astronomy and Astrophysics, 5071 West Saanich Road, Victoria, BC V9E 2E7, Canada\\
$^{12}$ Laboratoire Lagrange, Universit\'e de Nice Sophia-Antipolis, Observatoire de la C\^ote d'Azur, CNRS, \\
             Bd de l'Observatoire, CS 34229, 06304 Nice cedex 4, France\\
$^{13}$ GEPI, Observatoire de Paris, PSL Research University, CNRS, Place Jules Janssen, 92190 Meudon, France\\
$^{14}$ Laboratoire d'astrophysique, \'Ecole Polytechnique F\'ed\'erale de Lausanne (EPFL), Observatoire, 1290 Versoix, Switzerland\\
$^{15}$ Dept. of Physics and Astronomy, University of Victoria, P.O. Box 3055, STN CSC, Victoria BC V8W 3P6, Canada\\
$^{16}$ STFC UK Astronomy Technology Centre, Royal Observatory, Blackford Hill, Edinburgh, EH9 3HJ, UK \\
$^{17}$ Kapteyn Astronomical Institute, University of Groningen, Landleven 12, 9747AD Groningen, Netherlands\\
}

\date{\today}
\begin{document} 
\maketitle 
\begin{abstract} 
We present a detailed study of the faint Milky Way satellite Draco II (Dra II) from deep CFHT/MegaCam broadband $g$ and $i$ photometry and narrow-band metallicity-sensitive CaHK observations, along with follow-up  Keck II/DEIMOS multi-object spectroscopy. Forward modeling of the deep photometry allows us to refine the structural and photometric properties of Dra II: the distribution of stars in colour-magnitude space implies Dra II is old (13.5 $\pm 0.5$ Gyr), very metal poor, very faint ($L_V = 180 ^{+124}_{-72} \lsun$), and at a distance $d = 21.5 \pm 0.4 \kpc$. The narrow-band, metallicity-sensitive CaHK Pristine photometry confirms this very low metallicity ($\FeH = -2.7 \pm 0.1$ dex). Even though our study benefits from a doubling of the spectroscopic sample size compared to previous investigations, the velocity dispersion of the system is still only marginally resolved ($\sigma_{vr}<5.9 \kms$ at the 95 per cent confidence level) and confirms that Dra II is a dynamically cold stellar system with a large recessional velocity ($\langle v_{r}\rangle  = -342.5^{+1.1}_{-1.2}\kms$). We further show that the spectroscopically confirmed members of Dra~II have a mean proper motion of $(\mu_\alpha^*,\mu_\delta)=(1.26 \pm 0.27,0.94 \pm 0.28) \masyr$ in the Gaia DR2 data, which translates to an orbit with a pericenter and an apocenter of $21.3 ^{+0.7}_{-1.0}$ and $153.8 ^{+56.7}_{-34.7} \kpc$, respectively. Taken altogether, these properties favour the scenario of Dra~II being a potentially disrupting dwarf galaxy. The low-significance extra-tidal features we map around the satellite tentatively support this scenario.

\end{abstract}
 

\section{Introduction}
During the last decades, important photometric surveys such as the Sloan Digital Sky Survey \citep[SDSS]{york2000}, the Panoramic Survey Telescope And Rapid Response System, Pan-STARRS1 \citep[PS1]{chambers16},  or the Dark Energy Survey \citep[DES]{abbott05} have led to the discovery of dozens of Milky Way satellites. Some of these systems are extremely faint \citep[e.g.,][]{belokurov07,bechtol15,drlica-wagner15,kim15,koposov15,laevens15,martin15}, but studying them is important in order to better constrain the low-mass end of the galaxy mass function \citep{koposov09}. Moreover, systems confirmed to be dwarf galaxies are thought to be among the most dark matter dominated systems in the universe, potentially making them one of the best locations to test the standard cosmological model $\Lambda$CDM \citep[e.g.,][]{bullock17}.

However, the distinction between dwarf galaxies and globular clusters can be challenging \citep[e.g.][]{willman_strader12,laevens14} yet crucial. In the $\Lambda$CDM model, dwarf galaxies are located in massive dark matter halos. Thus, they have deep potential wells that can leave a trail of indirect observational evidence. For instance, they are more extended for a given luminosity, which explains the low surface-brightness nature of those systems and why deep photometric surveys were needed to reveal their existence. Dwarf galaxies are overall dynamically hot (i.e. their velocity dispersion is larger than that implied by the mass stored in their baryons alone, \citealt[\eg][]{martin07,simon07}), thus implying the presence of a much higher mass than can be estimated from their stars alone, while the typical velocity dispersion for faint clusters is of order tenths of $\kms$. Dwarf galaxies also share a few chemical properties: they are overall more metal-poor than old globular clusters with the same luminosity, and show evidence of a large metallicity spread, which indicates that the system has undergone chemical enrichment \citep{willman_strader12,kirby13}. This is a strong indirect evidence for the presence of a dark matter halo as the deeper potential well of dwarf galaxies allows them to retain their gas more efficiently against supernovae winds and shields them against reionization, therefore allowing for the formation of successive stellar populations through time, despite early star formation truncation \citep[\eg][]{brown14}. On the contrary, most Milky Way globular clusters show very low metallicity dispersion with $\sigma_\mathrm{[Fe/H]} < 0.1$ (\citealt{willman_strader12} and references therein). The few clusters with significant enrichment, such as $\omega$ Cen, are massive systems and even thought to be dwarf galaxy remnants \citep{bellazzini08,carretta10_cen}.

As the detection of fainter satellites enabled by deeper and deeper surveys continues, the line between dwarf galaxies and globular clusters becomes blurred. For this reason, dwarf galaxy candidates have to be studied thoroughly: deep observations in both photometry and spectroscopy are needed to constrain  the main chemical and structural properties of a given system.

Draco II (Dra~II) is a Milky Way satellite discovered by \citet{laevens15} in the Pan-STARRS1 3$\pi$ survey. At the time of its discovery, the satellite was found to be compact (half-light radius $r_h=19^{+8}_{-6} \pc$). \citet{martin16_dra} carried out the spectroscopic follow-up of Dra~II and inferred a marginally resolved velocity dispersion of $\sigma_{vr} = 2.9\pm2.1\kms$. Visual comparison of spectra of the few brightest Dra~II member stars suggested that the satellite could be metal poor ($\FeH < -2.1$) and could exhibit a metallicity spread. \citet{martin16_dra} tentatively favoured Dra~II being a dwarf galaxy, but pointed out that the velocity dispersion of the system is only marginally resolved. Furthermore, no bright giant stars ($g  < 19$) were identified as Dra~II members, making the estimate of the chemical properties of the satellite challenging. Due to the particular faintness of the satellite, and the small number of bright members, kinematic evidence for a DM halo was limited.

In this work, we reanalyze Dra~II and present a detailed study of its properties based on deep photometric observations obtained with the Megacam wide-field imager on the Canada-France-Hawaii Telescope (CFHT, \citealt{boulade03}) and Keck~II/DEIMOS spectroscopy \citep{faber03} that complements the sample of \citet{martin16_dra}. In particular, we include here novel narrow-band photometry that focuses on the metallicity-sensitive CaHK doublet. We use these observations, which are part of a specific dwarf-galaxy programme within the larger Pristine survey \citep{starkenburg17}, to identify the metal-poor Dra~II stars and estimate the metallicity and metallicity dispersion of the system.

The paper is arranged as follows: Section 2 describes the observations and data of both our photometry and spectroscopy; Section~3 focuses on the analysis of the deep broadband $g$ and $i$ photometry to infer the structural and photometric properties of Dra~II; Section~4 specifically centers on the study of the narrow-band CaHK observations to derive the metallicity and metallicity dispersion of the system; and Section~5 revises the multi-object spectroscopic study of Dra~II. The paper concludes with a discussion and conclusions in Section~6.

\section{Observations and Data}
\subsection{Photometry}
The photometry used in this paper was observed with the wide-field imager MegaCam on CFHT. It consists of deep, broadband observations with $g_\mathrm{MC}$ (487 nm) and $i_\mathrm{MC}$ (770 nm) Megacam filters and narrow-band observations with the new narrow-band CaHK Pristine filter that focuses on the metallicity-sensitive Calcium H\&K lines. This is the same filter that is used by the Pristine survey \citep{starkenburg17} to build a metallicity map of the Milky Way halo and search for the most metal-poor stars in the Galaxy. The data for Dra~II, which were observed before the official start of the Pristine survey, are now folded into a dedicated effort by the Pristine collaboration to observe all northern, faint Milky-Way dwarf galaxies or dwarf-galaxy candidates with this filter (the Pristine dwarf-galaxy survey).

Observations were conducted in service mode by the CFHT staff during the night of April 5th, 2016 during conditions of good seeing ($\sim0.5$--$0.7'')$. Multiple sub-exposures were observed in each band to better address CCD defects and facilitate cosmic ray removal. Exposure times amounted to $3\times700$\,s, $5\times500$\,s, and $3\times705$\,s in the $g_\mathrm{MC}$, $i_\mathrm{MC}$, and CaHK bands, respectively. After retrieval from the CFHT archive, the images are processed with a version of the Cambridge Astronomical Survey Unit pipeline \citep{irwin01}, which is specifically tailored to MegaCam data. We refer the reader to \citet{ibata14} for more details. The astrometric solution is derived using the catalogue of Pan-STARRS1 stars (PS1; \citealt{chambers16}) that are located in the field and have uncertainties on the $g_\mathrm{P1}$ PS1 photometry lower than 0.1 mag. The astrometric solution is good at the $\sim0.1''$ level.

MegaCam $g_\mathrm{MC}$ and $i_\mathrm{MC}$ bands are then transformed onto the PS1 photometric system by using the PS1 $g_\mathrm{P1}$ and $i_\mathrm{P1}$ catalogs. Unsaturated MegaCam point sources are cross identified with PS1 sources having photometric uncertainties below 0.05 mag. To derive the colour equations between the instrumental and PS1 magnitudes, we performed a second-order polynomial fit. We find

\begin{eqnarray}
g_\mathrm{MC} - g_\mathrm{P1} &  = & a^g_0x^{2}  + a^g_1x + a^g_2 ,\nonumber\\
i_\mathrm{MC} - i_\mathrm{P1} & = & a^i_0x^{2}  + a^i_1x + a^i_2,\nonumber
\end{eqnarray}

\noindent with $x \equiv g_\mathrm{MC} - i_\mathrm{MC}$. The calibration yields $a^g_0 = -0.0208 \pm 0.0021$, $a^g_1 = 0.0626 \pm 0.0051$, $a^g_2 = 3.5304 \pm 0.0052$ for the $g$ band and $a^i_0 = -0.0235 \pm 0.0019$, $a^i_1 = -0.0235 \pm 0.0048$, $a^i_2 = 4.2369 \pm 0.0047$ for the $i$ band. The uncertainties on the polynomials coefficients are propagated into the photometric uncertainties. For clarity, we drop the P1 subscripts in the rest of the text.

The narrow-band CaHK photometry is processed following the treatment presented in the paper describing the Pristine survey and includes specific calibration steps to deal with variations in the photometry as a function of the position in the field of view \citep{starkenburg17}. The Pristine model that translates ($CaHK$, $g$, $i$) into $\FeH$ is recalculated for the PS1 photometric system and applied to the Draco~II photometry.

All MegaCam magnitudes are dereddened following \citet{schlegel98} and using the extinction coefficients from \citet{schlafly11}, but it is worth noting that Dra~II is located in a low extinction area of the sky, with a median $E(B-V)$ of 0.018 mag. We rely on the CASU flags to isolate point sources. The MegaCam photometry are deeper than the original PS1 photometry that enabled the discovery of Dra~II but this means that the MegaCam data saturate for magnitudes brighter than $i  \sim 17.7$. For this reason, we complement the MegaCam data set with the PS1 photometry for magnitudes brighter than this limit. Finally, we clean the sample from stars for which the information on either of the two broadbands is missing, we discard stars with photometric uncertainties larger than 0.2 mag in either of the two bands, and we further discard faint sources with  $g < 24.5$. This latter cut removes regions of the colour-magnitude diagram (CMD) for which the star/galaxy separation becomes inefficient and the data are contaminated by a large number of background compact galaxies.

\begin{figure*}
\begin{center}
\centerline{\includegraphics[width=\hsize]{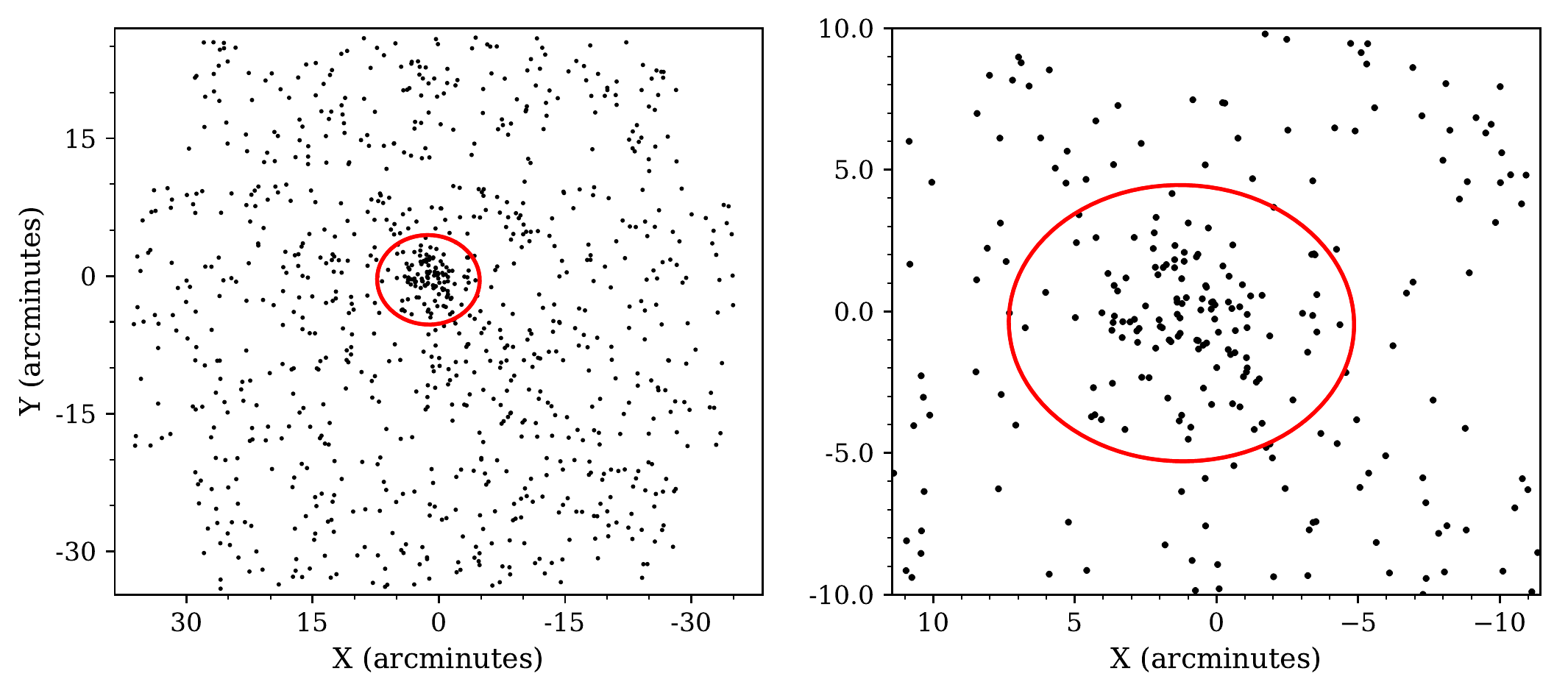}}
\caption{{\textit{Left panel: }}Distribution of MegaCam stars corresponding to a Dra~II-like stellar population, centred on the system. The mask selecting Dra~II-like stars only is shown in Figure 2. The red line represents the two half-light radii ($r_h \sim 3.0'$, $\epsilon \sim 0.23$) region of Dra~II based on the favoured model found in section 3. {\textit{Right panel: } Magnified view of the central region.} }
\label{radec_shape} 
\end{center}
\end{figure*}

The final photometric sample comprises 12,638 stars with broadband photometry, out of which 3,238 also have good quality $CaHK$ magnitudes. The spacial distribution of a fraction of this sample, composed only of Dra~II-like stars, is shown in  the left panel of Figure~\ref{radec_shape}, where the system is clearly visible as a compact stellar overdensity.

\subsection{Spectroscopy}

Dra~II was observed during two different runs using the Deep Extragalactic Imaging Multi-Object Spectrograph (DEIMOS) in multi-object spectroscopy mode: a first mask was observed in 2015 and was the focus of the study presented by \citet{martin16_dra} whilst the second run was observed a year later on September 4th, 2016. We used our group's standard set-up for these observations, employing the OG550 filter, the 1200~lines~mm$^{-1}$ grating and a central wavelength of $7800$\AA. This results in a FWHM resolution at our central wavelength of $\sim1.3$\AA, and covers a wavelength range of $\sim6500-9000$\AA. Such a setup allows us to well-resolve the Ca~II triplet lines at $\sim 8500$\AA. These strong absorption features are used to measure the line-of-sight velocities of our observed stars. The mask was observed for $1$ hour, split into $3\times1200$ seconds exposures. 

Stars were selected for targeting using the colour-magnitude diagram for Dra~II and they were given a priority for observation based on their distance from a fiducial isochrone, which highlighted the main sequence turn-off (MSTO), sub-giant and red giant branch of Dra~II. We then designed a slitmask using the IRAF DSIMULATOR software package provided by Keck Observatories. In total, 96 stars were selected for observation, and 73 of these targets returned spectra of sufficiently high $S/N$ such that a velocity could be measured using the pipeline detailed in \citet{ibata11} and \citet{martin16_dra}.  All stars with a signal-over-noise ratio below 3.0 or a velocity uncertainty greater than 15  km s$^{-1}$ were finally discarded. Heliocentric velocities and equivalent widths from stars observed twice are transformed into one single measurement by computing the weighted mean and uncertainties from the two independent velocity measurements. We do not investigate the potential binarity of Dra~II stars in great detail as the low signal-over-noise of the spectra translate into typical velocity uncertainties in the range 5-15 $\kms$, which can make the detection of any variability challenging. 
\citet{martin16_dra} presented a spectroscopic study of Dra~II using the 2015 dataset, however, the heliocentric velocities of the 2015 stars in this work are slightly different: using the usual method of the \citet{ibata11} pipeline to derive the velocities, the average difference of the 2015 and 2016 velocities is not $0 \kms$ as expected, but is shifted of a few $\kms$. These effects appear when the velocities are derived through a non-flexible, but supposedly more precise method in the pipeline of \citet{ibata11}, that was used in the paper of \citet{martin16_dra}. In this work, using a slightly less precise, but more flexible method of the same pipeline to extract the velocities, we are able to get rid of these systematics and find the expected mean difference in velocities for all stars observed twice of 0 $\kms$.

\section{Broadband photometry analysis}

\begin{figure*}
\begin{center}
\centerline{\includegraphics[scale=0.55]{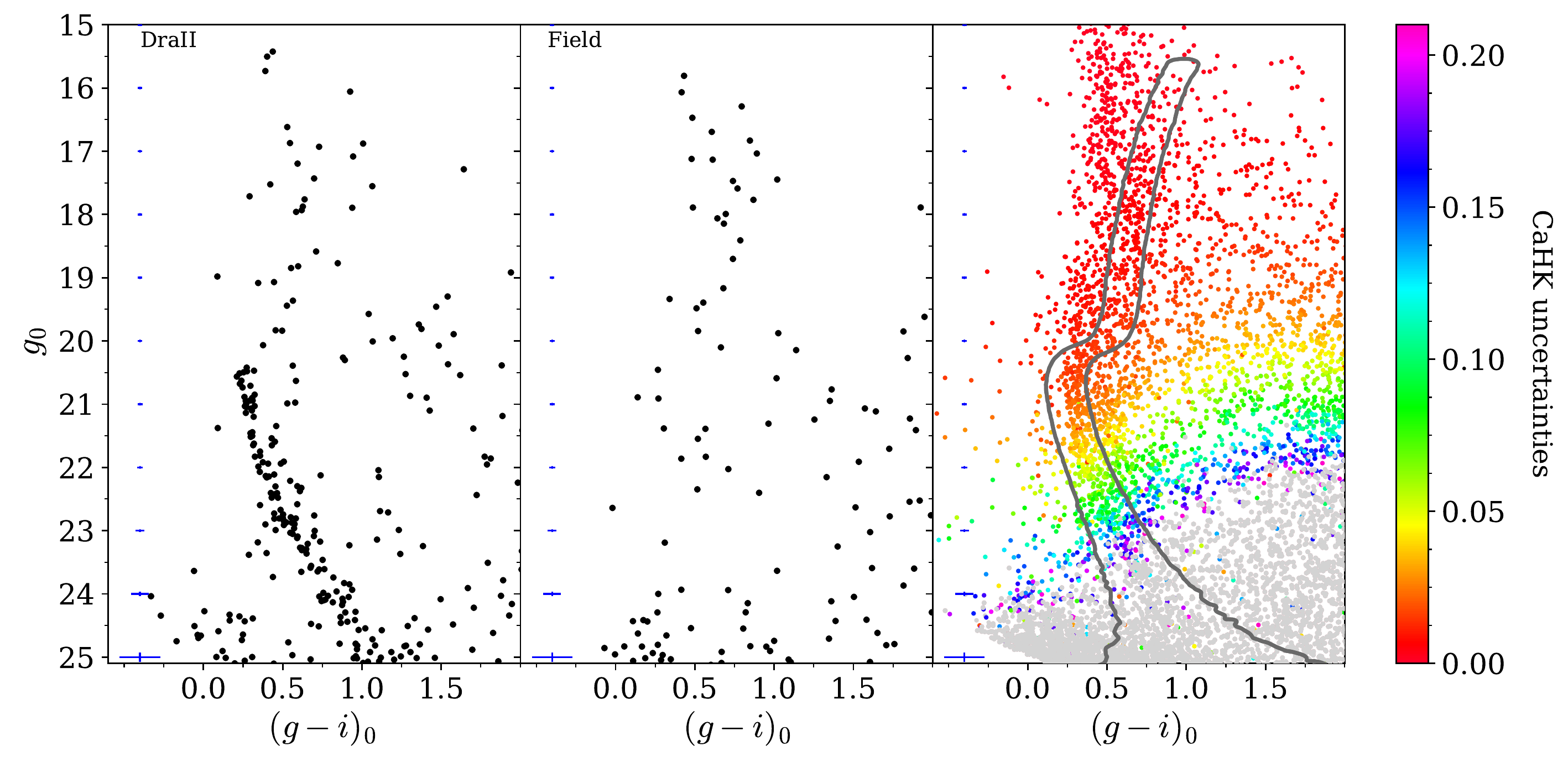}}
\caption{\textit{Left panel:} CMD of stars within two half-light radii ($r_h \sim 3.0'$) of the Dra II centroid. The main sequence of Dra~II clearly stands out and points towards an old and metal-poor stellar population. The satellite seems to have very few, if any, giant stars. The photometric uncertainties in the $g$ band and $g-i$ colour are shown every magnitude on the left of each panel. \textit{Middle panel:} The field CMD obtained within a similarly sized region $\sim$ 25 arcminutes away from Dra~II centroid. \textit{Right panel:} CMD of all stars in the photometric dataset, colour-coded according to the $CaHK$ photometric uncertainties. Stars coloured in grey have $CaHK$ uncertainties above 0.2. The CaHK is clearly shallower than the broadband $g$ and $i$ photometry. Finally, the mask selecting only Dra~II-like stellar population is shown in solid, dark-grey line. }
\label{CMDs}
\end{center}
\end{figure*}

The CMD of Dra~II for sources within two half-light radii ($2r_h$, see below) is presented in Figure~\ref{CMDs} (left panel). For comparison, the CMD of a field region of the same coverage but selected in the outskirts of the MegaCam field of view is shown in the middle-left panel. The main sequence observed in the Dra II CMD is consistent with an old and metal-poor stellar population (see below) as originally pointed out by \citet{laevens15}, but the MegaCam data is much deeper and traces the main sequence of the system more than three magnitudes below the turnoff. The exquisite MegaCam CMD is highlighted by the narrowness of this sequence. The 50 per cent completeness of the data in the $g$ band is reached at $g = 25.2$ mag and $i = 23.9$ mag. We confirm that the main sequence of Dra~II contains very few stars brighter than the turnoff and that the satellite is particularly faint. Anticipating on the spectroscopic analysis presented below, the right panel of Figure~\ref{spectro_plot} highlights stars with radial velocity measurements. Likely Dra~II members appear in red with $v_r\sim-345\kms$. With these velocities, it is possible to isolate a handful of potential Dra~II stars just above the turnoff. We find no bright RGB stars and no horizontal branch stars in the system.

\subsection{Structural and CMD analysis}

We take advantage of the deep MegaCam data and of the better sampling of the system to revisit the structural analysis performed by \citet{laevens15}. The  analysis is based heavily on the algorithm presented in \citet{martin08} and \citet{martin16} and we separately infer the CMD-properties of Dra~II. Altogether, we aim to estimate the structural properties of the system (the coordinate offsets of the centroid from the literature values, $X_{0}$ and $Y_{0}$, the half-light radius along the major axis, $r_h$, the ellipticity\footnote{The ellipticity is defined as $\epsilon = 1 - \frac{a}{b}$, with $a$ and $b$ the major and minor axes of the ellipse respectively.}, $\epsilon$, the position angle of the major axis east of north, $\theta$, and the number of stars within the MegaCam data, $N^*$), along with its distance modulus $m-M$, Age $A$, metallicity $\FeH_\mathrm{CMD}$, and abundance in $\alpha$ elements $[\alpha/\textrm{Fe}]$.

For any star $k$ in our sample, the pieces of information used at this stage are the coordinates of the star $X_k$ and $Y_k$, projected on the sky on the plane tangent to Dra~II's centroid, and the MegaCam magnitudes, $g_k$ and $i_k$. For clarity, we define $\vec{d_{k}}^{sp} \equiv \{ X_{k},Y_{k} \}$ and $\vec{d_{k}}^{CMD} \equiv \{ g_{k},i_{k} \}$. The suite of parameters we aim to infer is divided into a set of structural parameters $\mathcal{P}_\mathrm{sp} \equiv \{X_{0},Y_{0}, r_{h}, \epsilon, \theta, N^*, \eta_\mathrm{sp}\}$ and a set of CMD-related parameters, $\mathcal{P}_\mathrm{CMD} \equiv \{ m-M,  A, \FeH_\mathrm{CMD}, [\alpha/Fe], \eta_\mathrm{CMD}\}$, with $\eta_\mathrm{sp}$ and $\eta_\mathrm{CMD}$ the fractions of Dra~II stars in the spacial and CMD data sets. Following these definitions and keeping in mind that any star could be a Dra~II star or a field star that belongs to the Milky-Way contamination, we can express the spacial likelihood of star $k$ as

\begin{equation}
\ell^{tot}_\mathrm{sp}(\vec{d}_{k,sp} | \mathcal{P_\mathrm{sp}}) = \eta_\mathrm{sp} \ell^\mathrm{DraII}_\mathrm{sp}(\vec{d}_{k,sp} | \mathcal{P_\mathrm{sp}}) + (1 - \eta_\mathrm{sp} ) \ell^\mathrm{MW}_\mathrm{sp}(\vec{d}_{k,sp}),
\end{equation} 

\noindent where $\ell^\mathrm{DraII}_\mathrm{sp}$ and $\ell^\mathrm{MW}_\mathrm{sp}$ are the spacial likelihoods of star $k$ in the Dra~II or the field-contamination models, respectively.

We follow \citet{martin16} and assume that Dra~II stars follow an exponential radial density profile whereas the field contamination is taken to be flat over the MegaCam field of view. However, and contrary to \citet{martin16}, we assume $N^*$ is a parameter to be determined by the normalization. For this reason, the formalism is slightly different here. The radial density profile of the system is expressed as

\begin{equation}
    \rho_\mathrm{dwarf}(r) = \frac{1.68^{2}}{2 \pi r_h^2(1-\epsilon)}\exp(-1.68\frac{r}{r_h}),
\end{equation} 

\noindent with $r$ the elliptical radius, which relates to projected sky coordinates $(x,y)$ via

\begin{eqnarray}
\label{eqn:r}
r=\bigg[  \Big(\frac{1}{1-\epsilon} ((X-X_0)\cos\theta - (Y-Y_0)\sin\theta)\Big)^2\nonumber\\
+ \Big((X-X_0)\sin\theta +(Y-Y_0)\cos\theta\Big)^2  \bigg]^{1/2}.
\end{eqnarray}

The spacial likelihood of the Dra~II component of the model is then simply

\begin{equation}
\ell_\mathrm{sp}^\mathrm{DraII}(X_k,Y_k)  =  \frac{\rho_\mathrm{dwarf}(r)}{\int_{\mathcal{A}} \rho_\mathrm{dwarf}(r) d\mathcal{A} },
\end{equation}

\noindent where $\mathcal{A}$ is the area of the sky over which the analysis is conducted.

The spacial likelihood of the Milky Way contamination model is much simpler and, with our assumption that it is constant, we simply have

\begin{equation}
\ell_\mathrm{sp}^\mathrm{MW}  =   \frac{1}{\int_{d\mathcal{A}}}.
\end{equation} 

\begin{figure}
\begin{center}
\centerline{\includegraphics[scale=0.4]{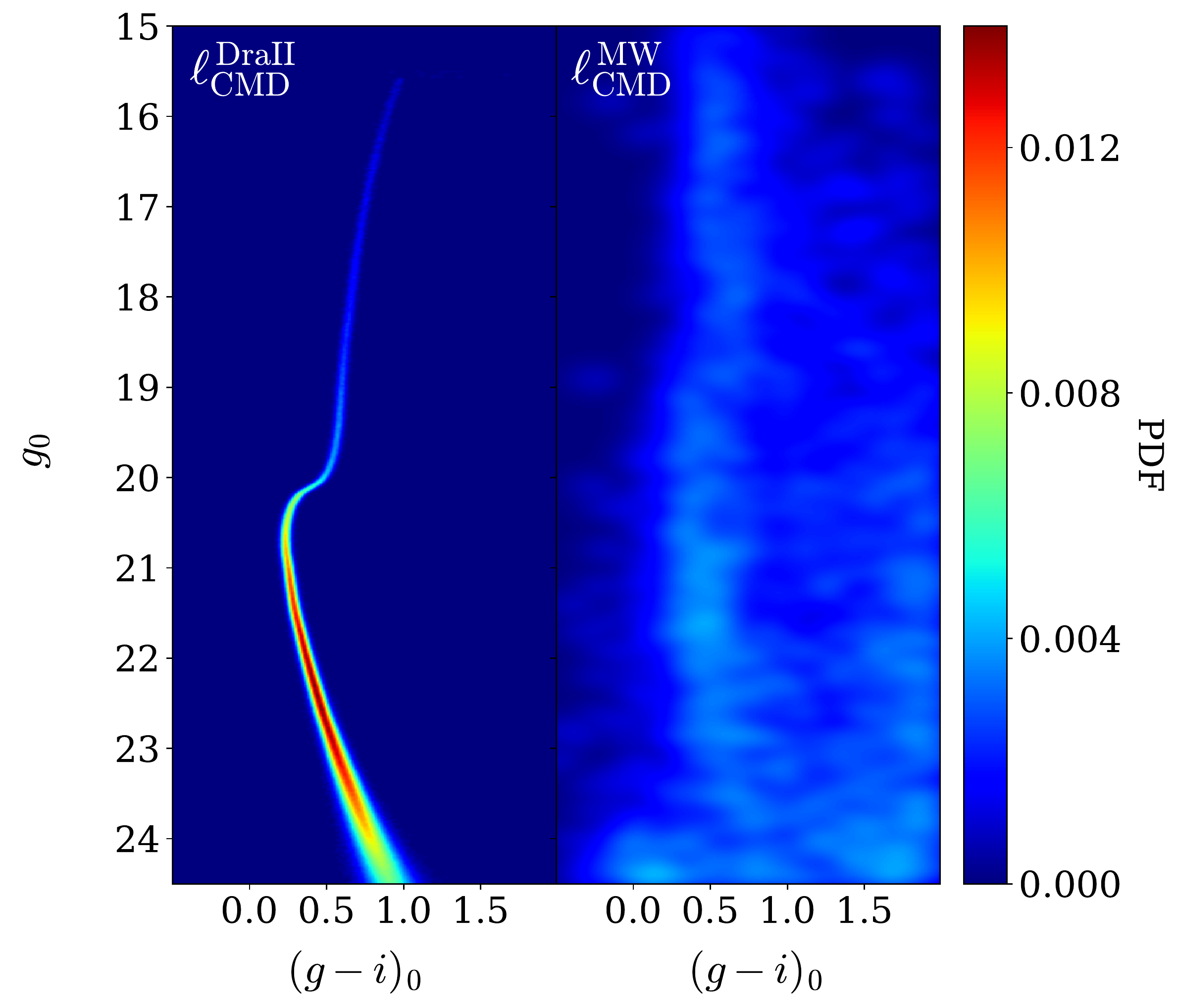}}
\vspace{-1em}
\caption{{\textit{ Left panel }}: Likelihood of the stellar population favoured by our model ($A$: 13.5 Gyr, $\FeH_\mathrm{CMD} = -2.4$, [$\alpha$/Fe] = 0.6 and a distance modulus of 16.67). It is constructed from a theoretical isochrone weighted by its luminosity function, convolved by the typical Megacam photometric uncertainties, and finally weighted by the compltenesses in $g_0$ and $i_0$.  {\textit{ Right panel: }} Likelihood of the contamination stars. The maximum density for the left panel is far greater than for the background likelihoood on the right so the two are represented with a square-root colour scale.}
\label{proba_mw} 
\end{center}
\end{figure}

Similarly, one can express the total CMD likelihood as followed :

\begin{eqnarray}
\ell^{tot}_\mathrm{CMD}(\vec{d}_{k,CMD} | \mathcal{P_\mathrm{CMD}}) = \eta_\mathrm{CMD} \ell^\mathrm{DraII}_\mathrm{CMD}
(\vec{d}_{k,CMD} | \mathcal{P_\mathrm{CMD}}) \nonumber \\  
+ (1 - \eta_\mathrm{CMD} ) \ell^\mathrm{MW}_\mathrm{CMD}(\vec{d}_{k,CMD}),
\end{eqnarray} 

\noindent where $\ell^\mathrm{DraII}_\mathrm{CMD}$ and $\ell^\mathrm{MW}_\mathrm{CMD}$ are the CMD likelihoods of star $k$ in the Dra~II or the field-contamination models, respectively.

To build the CMD models, we rely on a set of isochrones for Dra~II and build an empirical model for the field contamination.  We base our CMD model of Dra~II, $\ell_\mathrm{CMD}^\mathrm{DraII}$, on a set of Dartmouth isochrones and luminosity functions\footnote{\url{http://stellar.dartmouth.edu/models/webtools.html}} \citep{dotter08} calculated for the PS1 photometric system. For a given set of CMD parameters $\mathcal{P}_\mathrm{CMD}$, we download the isochrone and luminosity function (LF) of the stellar population of this age $A$, metallicity $\FeH_\mathrm{CMD}$, and $\alpha$ abundance $[\alpha/\textrm{Fe}]$, and shift it by the distance modulus $m-M$. Since the isochrones and LFs provided by the Darmouth library are not continuous but discrete tracks, they are linearly splined. The isochrones are then weighted according to their associated LF. At this stage, each isochrone is a continuous track in CMD space with a `height' equal to the luminosity function along it. We then generate a CMD PDF of where the system stars are likely to be located by simply convolving this track with the photometric uncertainties. With this formalism, we implicitly assume that Dra~II contains a single stellar population and any intrinsic spread in the properties of the system will generate wider posterior PDFs. However, as isochrones pile up towards the blue in the metal-poor end regime ([Fe/H] $< -1.4$), only significant metallicity or age gradients would affect our results. Finally, the colour-magnitude space over which the PDF is calculated is implemented with pixel sizes of 0.01 magnitude on the side, so we further convolve the resulting PDF by a Gaussian of dispersion 0.01 mag to avoid aliasing issues in our representation of the PDF. Since this PDF is supposed to describe the observed Dra~II features of the CMD, the completeness of the data needs to be taken into account, therefore, each track is weighted by the product of the completenesses in $g_0$ and $i_0$. This completeness is computed following the model built by \citet{martin16} on similar MegaCam data, simply shifted to the appropriate reference median magnitude (the median magnitude of all stars in our photometry with photometric uncertainties between 0.09-0.11). The final step normalizes this PDF to unity so it is properly defined. An example of the resulting model is shown in the left-hand panel of Figure~\ref{proba_mw} for the specific set of parameters $\mathcal{P}_\mathrm{CMD}=\{m-M = 16.67, A = 13.5\Gyr, \FeH_\mathrm{CMD} = -2.4, [\alpha/\textrm{Fe}] = +0.6\}$.

The model for $\ell_\mathrm{CMD}^{MW}$ is built empirically from the CMD position of field stars in the MegaCam data. We select all stars beyond $5 r_h$ and bin them in CMD space. Each bin has a width of 0.01 mag along both the magnitude and the colour directions. In order to diminish the amount of shot noise in the resulting binned CMD, we further smooth it with a Gaussian kernel of width 0.1 mag in both dimensions. The resulting smoothed CMD is presented in Figure~\ref{proba_mw} after its normalization so it is a properly defined PDF.

With the model being entirely defined, we can now focus on the inference on the model's parameters. Since the structural side of the analysis can be biased by the presence of the chip gaps visible in Figure~\ref{radec_shape}, they are accounted for by constructing a binary mask correcting the effective area of the field. From the $N_\mathrm{tot}$ stars present in this region, the spacial likelihood $\mathcal{L_\mathrm{sp}}$ (resp. for the CMD) of a given model is 

\begin{equation}
\mathcal{L_\mathrm{sp}}\left(\{\vec{d}_{k,sp}\}|\mathcal{P_\mathrm{sp}}\right) = \prod_{k = 1}^{N_\mathrm{tot}} \ell_\mathrm{sp}^{tot}\left(\vec{d}_{k,sp} | \mathcal{P_\mathrm{sp}}\right)
\end{equation}

\noindent and the posterior probability we are after is, trivially,

\begin{equation}
P_\mathrm{sp}\left(\mathcal{P_\mathrm{sp}}|\{\vec{d}_{k,sp}\}\right) \propto \mathcal{L_\mathrm{sp}}\left(\{\vec{d}_{k,sp}\}|\mathcal{P_\mathrm{sp}}\right) P_\mathrm{sp}(\mathcal{P_\mathrm{sp}}),
\end{equation}

\noindent with $P_\mathrm{sp}(\mathcal{P_\mathrm{sp}})$ the combined prior on the model parameters. These priors are listed in Table~\ref{tbl-2} and are chosen to be uniform for an old stellar population, with distance and structural parameters loosely close to the favoured parameters according to \citet{laevens15}. Anticipating on section 4, the systemic metallicity of the satellite is found to be $\langle\FeH_\mathrm{DraII}^{\mathrm{CaHK}}\rangle = -2.7 \pm 0.1$\,dex using the narrow-band, CaHK photometry. This result is used as a Gaussian prior to the CMD analysis.

\begin{figure*}
\begin{center}
\centerline{\includegraphics[width=\hsize]{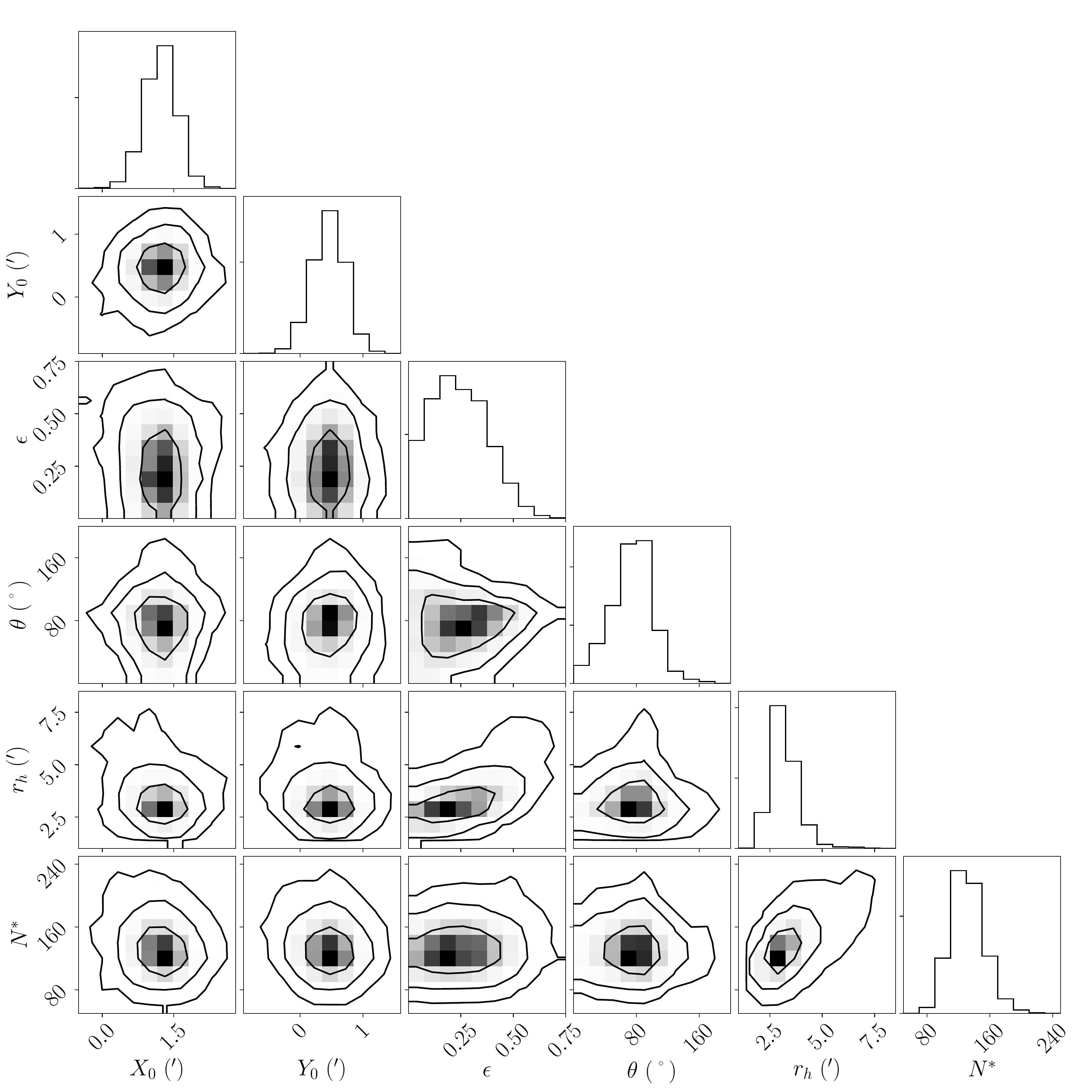}}
\caption{One- and two-dimensional posterior PDFs of the structural parameters of Dra~II, inferred using the method described in section 3.1. Contours correspond to the usual 1, 2 and 3$\sigma$ confidence intervals in the case of a two-dimensional Gaussian.}
\label{corner_sp}
\end{center}
\end{figure*}

\begin{figure*}
\begin{center}
\centerline{\includegraphics[width=\hsize]{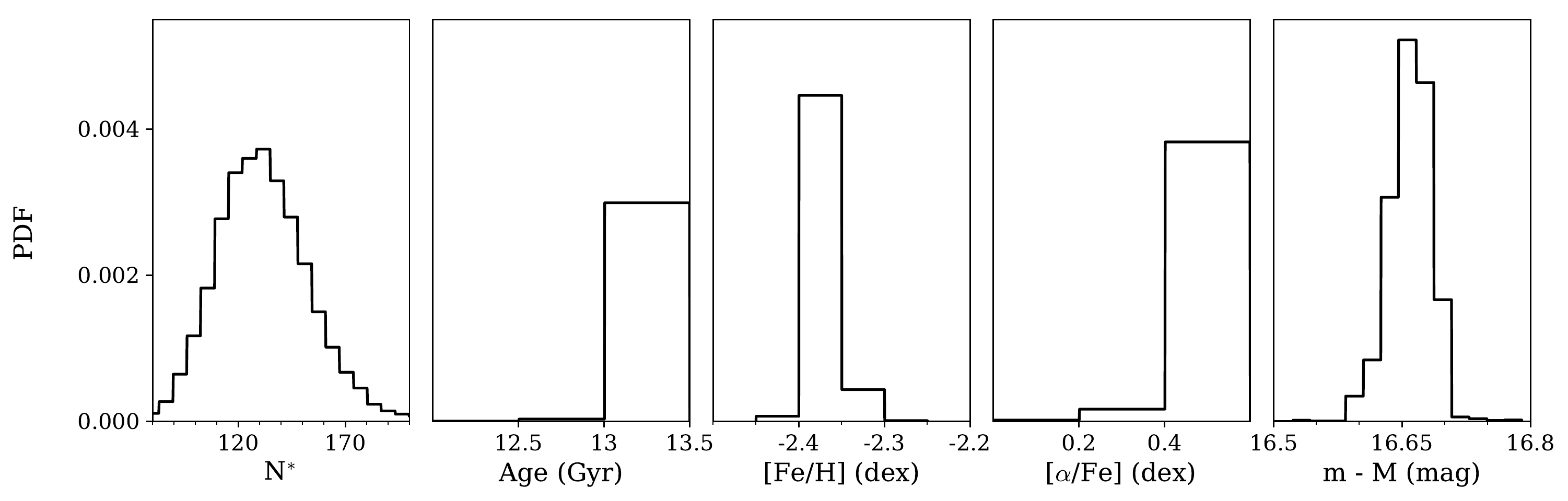}}
\caption{One-dimensional PDFs of the CMD parameters of Dra~II.}
\label{corner_CMD}
\end{center}
\end{figure*}

In order to build the posterior N-dimensional distribution function, we devised our own Markov Chain Monte Carlo code based on a Metropolis-Hastings algorithm \citep{hastings70}. To ensure convergence, we aim for an acceptance ratio of $\sim25$ per cent and run the algorithm for a few million iterations. Convergence is not an issue for this large number of iterations. Finally, for the CMD analysis, we restrict ourselves to a specific region of the CMD: a visual inspection of the Dra~II main sequence in Figure \ref{CMDs} shows that all stars outside $-0.5 < (g - i)_0 < 2.0$ are contaminants. For this reason, there is no need to take them into account in our analysis, and the following CMD and structural analyses are performed only with stars with $15 < g_0 < 24.5$, and $-0.5 < (g - i)_0 < 2.0$.  The resulting two-dimensional marginalised PDFs are presented in Figure~\ref{corner_sp} for spacial and \ref{corner_CMD} for CMD parameters.

\begin{table*}
\renewcommand{\arraystretch}{1.25}
\begin{center}
\caption{Inferred properties of Dra~II.\label{tbl-2}}
\begin{tabular}{lcccc}
\hline
Parameter &  Unit & Prior & Favoured model & Uncertainties  \\
\hline

RA $ \alpha $ & degrees & --- & $238.174$ & $\pm 0.005$ \\
DEC $ \delta $ & degrees & --- & $+64.579$  &  $\pm 0.006$ \\
$r_{h}$ & arcmin & $> 0$ & 3.0 & $^{+0.7}_{-0.5}$ \\
$r_{h}$ & pc & & $19.0$ & $^{+4.5}_{-2.6}$ \\
$\theta$ & degrees & [0,180] & $76$ & $^{+22}_{-32}$ \\
$\epsilon$ & - & $> 0$ & $0.23$ & $\pm 0.15$ \\
Distance modulus & mag & [16.3,17.1] & 16.67 & $\pm 0.05$  \\
Distance & kpc & & $21.5$ &  $ \pm 0.4$ \\
Age & Gyr & [10,13.5] & $13.5$ & $\pm 0.5$  \\
$\FeH$ & dex & --- & $-2.7$ & $\pm 0.1$  \\
$\sigma_\mathrm{\FeH}$ & dex & $> 0$ & Unresolved & $< 0.24$ dex at $95$\%  \\
$[\alpha/\textrm{Fe}]$ & dex & [0.0,0.6] & $0.6$ &  $> 0.4 $ at $89$\% \\
$L_V$ & L$_{\odot}$ & $> 0$ & $180$ & $^{+124}_{-72}$  \\
M$_V$ & mag & --- & $-0.8$ &  $^{+0.4}_{-1.0}$ \\
$\mu_{0}$ & mag arcsec$^{-2}$ & --- & $28.1$ & $\pm 0.7$  \\
$<v_r>$ & $\kms$ & --- & $-342.5$ & $^{+1.1}_{-1.2}$  \\
$<v_r>_{gsr}$ & $\kms$ & --- & $-172.0$ & $^{+1.1}_{-1.2}$  \\
$\sigma_{vr}$ & $\kms$ & $>$ 0 & Unresolved & $< 5.9 \kms$ at $95$\%   \\
$\mu_{\alpha}^{*}$ & mas.yr$^{-1}$  & --- & 0.54 & $\pm 0.27$   \\
$\mu_{\delta}$ & mas.yr$^{-1}$ &  --- & 0.94 & $\pm 0.28$   \\

\hline
\end{tabular}
\end{center}
\end{table*}

\begin{figure}
\begin{center}
\centerline{\includegraphics[scale=0.45]{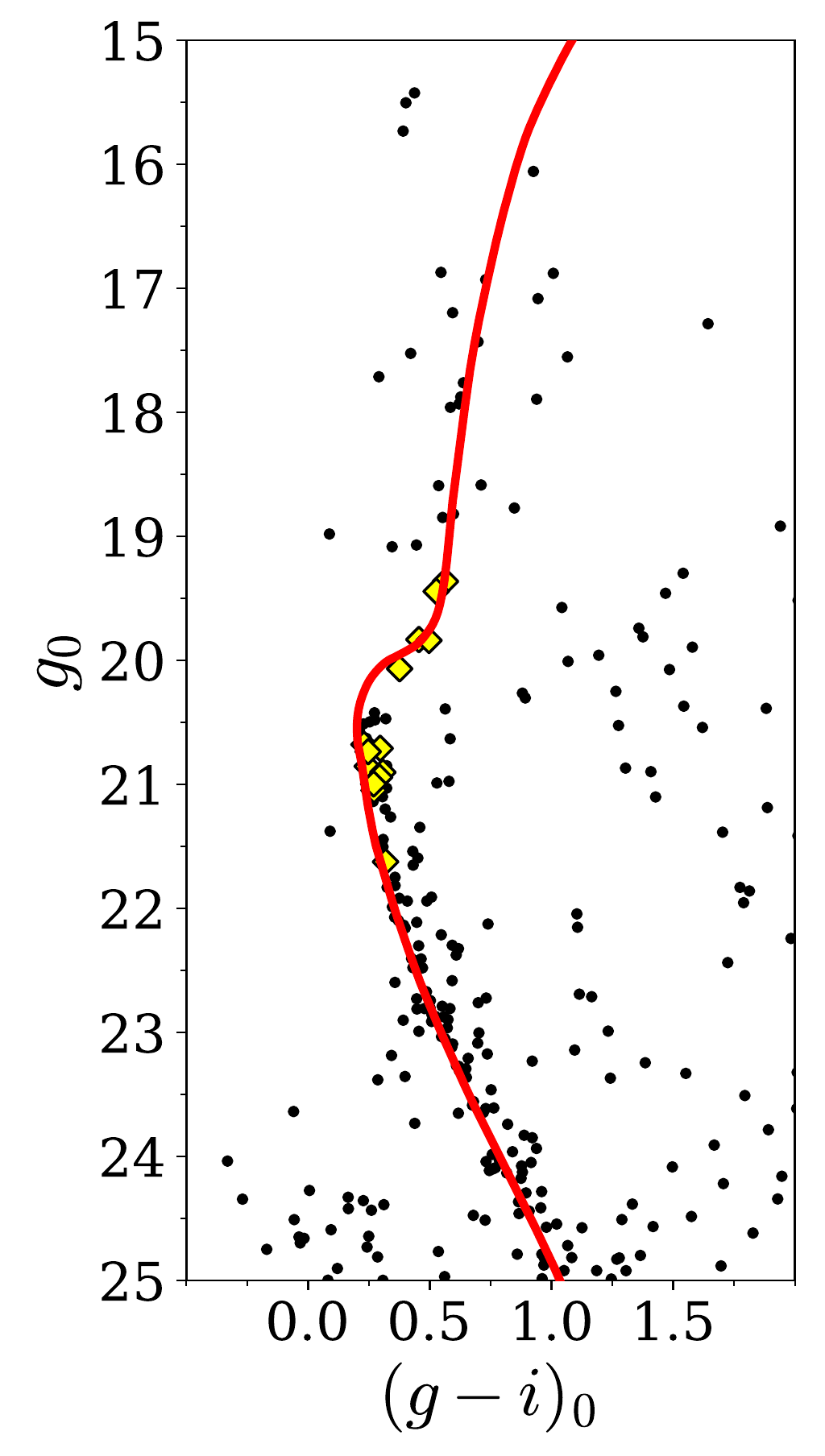}}
\caption{CMD of stars within $2r_h$ of Dra~II's centroid, along with the favoured isochrone found in section 3.1, corresponding to a stellar population of $13.5^{+0.5}_{-1.0} \Gyr$, $\FeH_\mathrm{CMD} = -2.40 \pm 0.05$ dex and $[\alpha/\textrm{Fe}] = +0.6$ dex. Stars confirmed as spectroscopic members in section 5 are represented as yellow diamonds.}
\label{best_fit_iso}
\end{center}
\end{figure}

\begin{figure}
\begin{center}
\centerline{\includegraphics[scale=0.45]{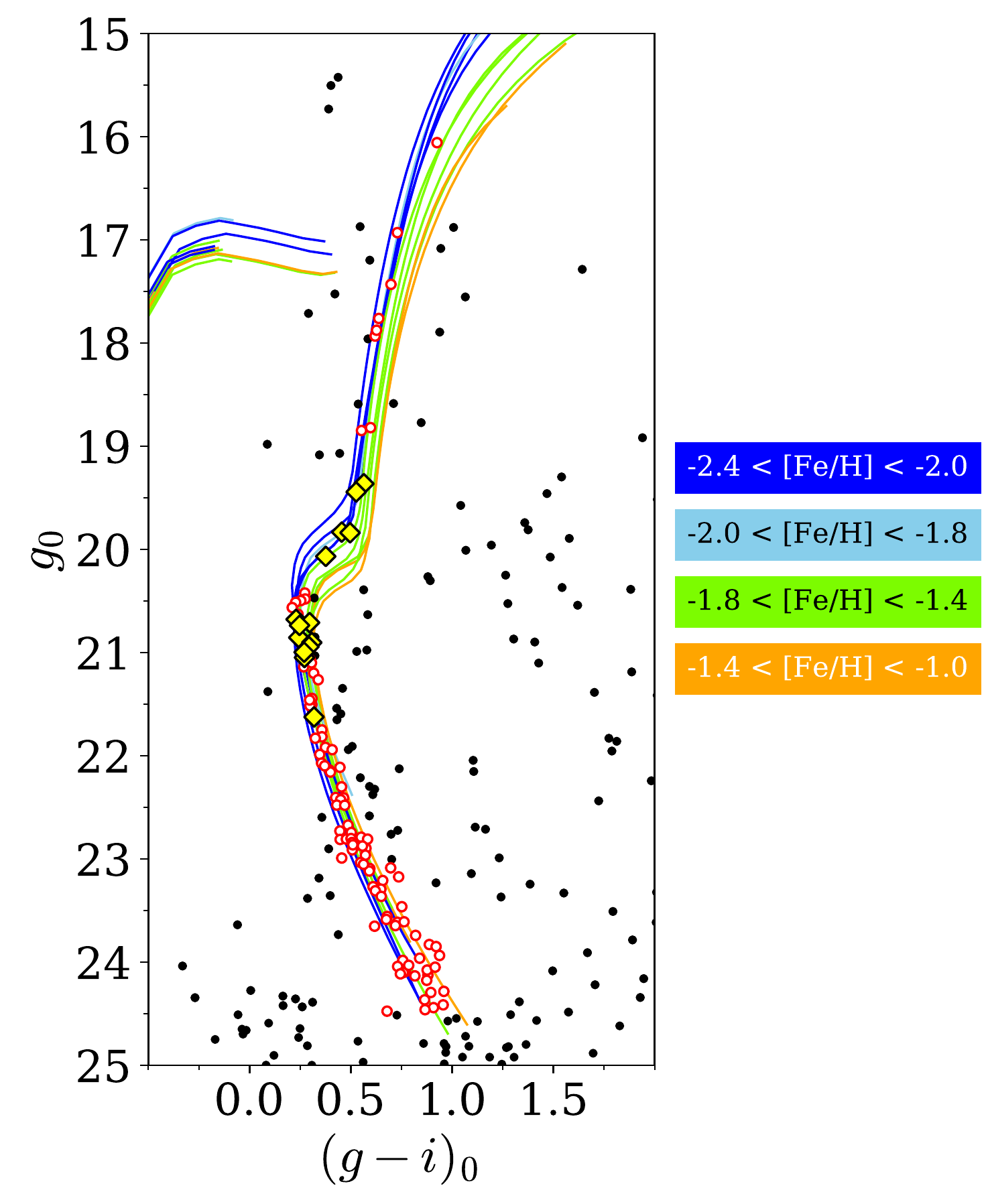}}
\caption{CMD of stars within two half-light radii of Dra II centroid. Several metal-poor globular cluster fiducials from \citet{bernard14} are represented and colour-coded by metallicity ranges. Red circled dots are stars with a Dra~II membership probability greater than 1 per cent. Yellow diamonds are Dra~II members confirmed by spectroscopy. The fiducials that best represent the Dra~II CMD features are the blue ones, with a metallicity range $-2.4 < [Fe/H] <-2.0$.}
\label{fid}
\end{center}
\end{figure}

\begin{figure}
\begin{center}
\centerline{\includegraphics[scale=0.45]{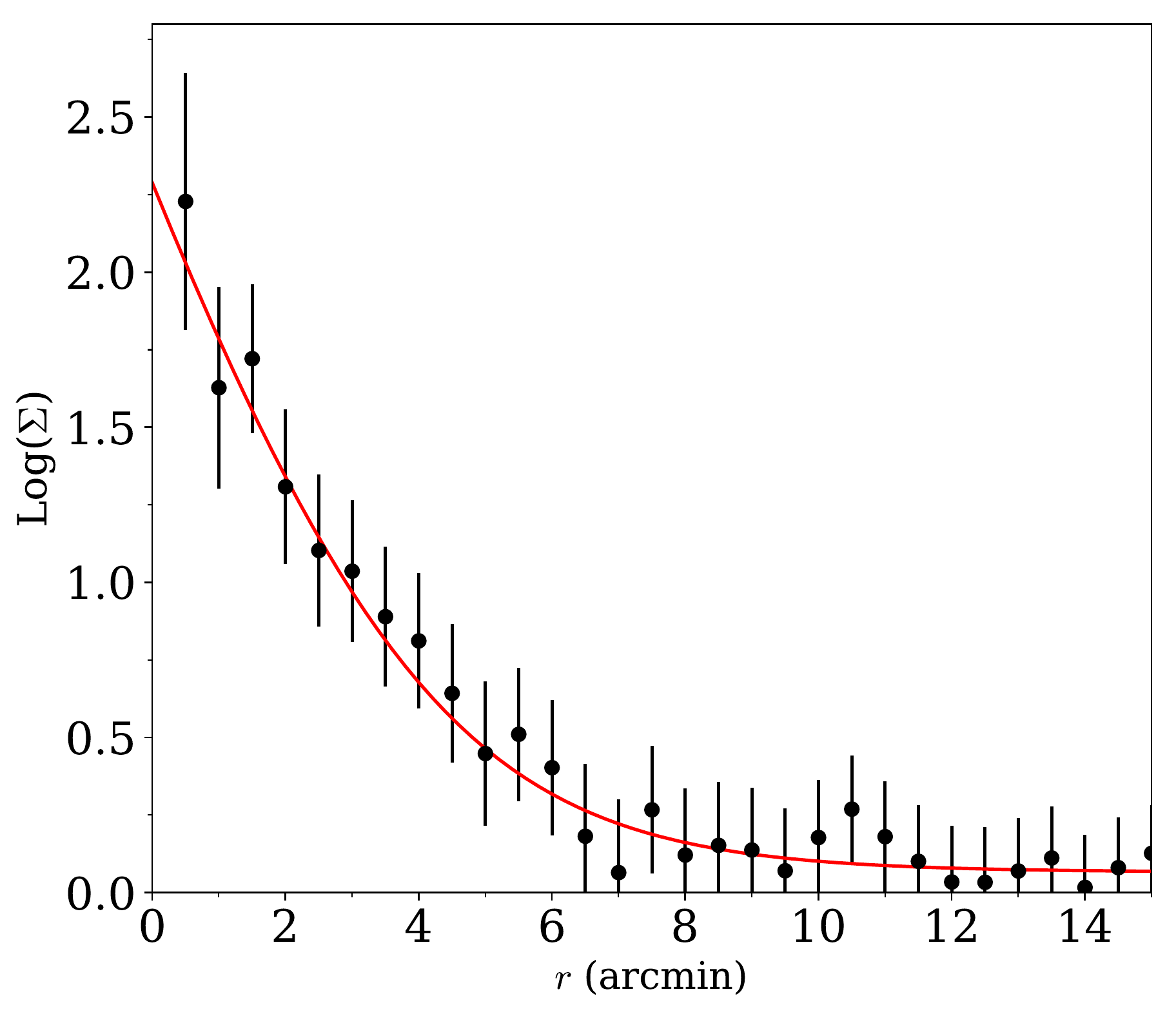}}
\caption{Comparison of the favoured exponential radial density profile (red solid line) with the binned data in elliptical annuli following the favoured structural model (dots). The error bars represents Poisson uncertainties on the number count of each annulus. $r$ is the elliptical radius.}
\label{radial_profile}
\end{center}
\end{figure}

Our results are  compatible with the ones presented by \citet{laevens15} in the discovery paper of Dra~II. From the deeper MegaCam data, we confirm the half-light radius of the satellite to be $r_h$ = $3.0^{+0.7}_{-0.5}$ arcmin (vs. $2.7^{+1.0}_{-0.8}$ arcmin before). Overall, the deeper MegaCam data allows for better constraints with smaller uncertainties on all parameters. The use of a Plummer profile instead of an exponential profile yields similar results. The radial profile of the favoured spacial model is presented in Figure \ref{radial_profile}.

The CMD part of the analysis yields a robust distance estimate ($m-M=16.67 \pm 0.05$ mag; or a heliocentric distance $d = 21.5 \pm 0.4 \kpc$) that is slightly smaller than the one proposed by \citet{laevens15}, who estimated a distance modulus of $\sim16.9$ by eye. The favoured isochrone also corresponds to a stellar population of $A=13.5 \pm 0.5 \Gyr $, $ \FeH_\mathrm{CMD} = -2.40 \pm 0.05$ dex, and [$\alpha$/Fe] = +0.6 dex. The 1D PDFs of the CMD parameters are shown in Figure \ref{corner_CMD}, while Figure \ref{best_fit_iso} shows that this stellar population is a good description of the features in the CMD of Dra~II and of the stars identified as members of the satellite through a spectroscopic study (see section 5). The choice of showing only 1D PDFs for the CMD inference is purely aesthetic: each parameter is chosen over a grid that can have large steps (e.g. [$\alpha$/Fe] is chosen over a grid with 0.2 dex step), which does not give representative or aesthetically pleasing 2D contours. It is however important to note that there is no clear correlation between the CMD parameters.

The alpha-abundance ratio of the favoured model is to be taken with caution as it reaches the limits of the [$\alpha$/Fe] range allowed by this set of isochrones. The alpha abundance of 0.6 found above is high but not totally unrealistic for a dwarf galaxy: \citet{vargas13} shows that faint Milky Way dwarf galaxies such as Segue 1 ($M_V \sim -1.5$) are compatible with this result.  Another fit was performed using an uniform prior in [$\alpha/Fe$] over the range [0.0,0.4] to test the analysis without reaching the end of the alpha abundance grid. This does not significantly change our results. 

 To investigate the impact of the choice of the completeness model used, the favoured CMD and spacial model are used to simulate a Dra~II-like population. The analysis is then performed three more times: one time with our actual completeness model, and two other times using the completeness model shifted by $\pm$ 0.5 mag respectively. The results of these analyses are all consistent within the uncertainties, showing that the impact of the completeness model is limited and that it does not significantly affect our results.

The systemic metallicity of the satellite appears consistent with the luminosity-metallicity relation for DGs and with the analysis previously proposed by \citet{martin16_dra} in a qualitative analysis of their spectra. We repeat the Calcium-triplet equivalent-width analysis of \citet{martin16_dra} for the three low-RGB stars with S/N $> 10$ that used the \citet{starkenburg10} relation. It is worth pointing out that this relation is calibrated for RGB stars. However, \citet{leaman13} implies that it can be applied to stars 2 magnitudes below the RGB and give consistent results. The analysis yields a systemic metallicity for Dra~II of $\FeH_\mathrm{spectro} = -2.43 ^{+0.41}_{-0.82}$ dex, which is compatible with our CMD analysis. Due to the lack of bright member stars (the brightest used in this analysis has $g_0 = 18.8$), it is challenging to obtain tight constrains on the spectroscopic metallicity of the satellite.

\begin{figure}
\begin{center}
\centerline{\includegraphics[scale=0.45]{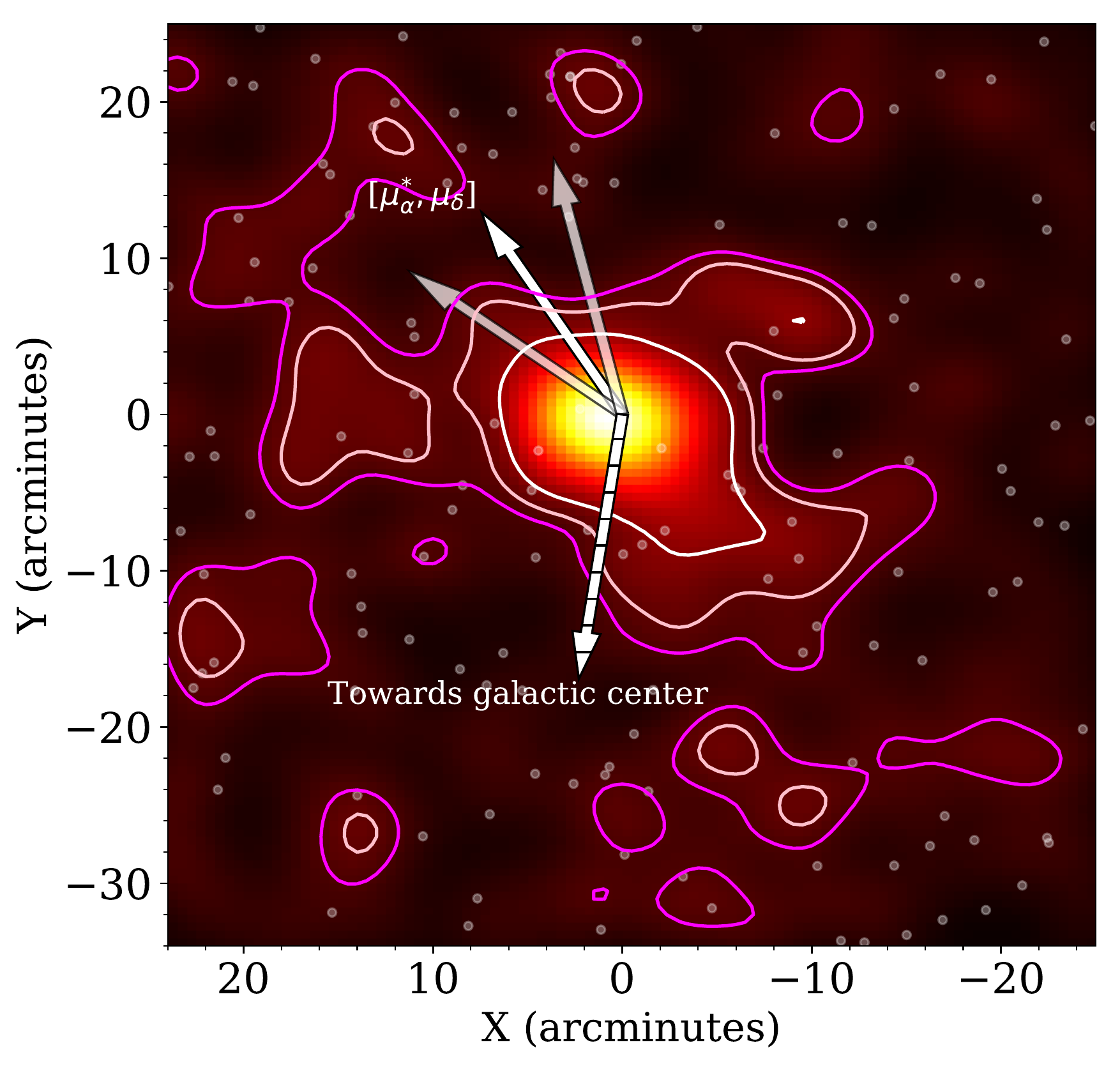}}
\caption{On-sky density plot of the full field of view for all stars with $\mathcal{P}_{mem} \geq 0.01$, smoothed using a 2 arcminutes gaussian kernel. Regions with a density within the upper 68, 95 and 99.85 per cent of the background pixels distribution are shown with magenta, pink, and white contours respectively. The dashed arrow shows the direction towards the Galactic center. The upper white arrow shows the favoured proper motion vector $\vec{\mu} =$ [$\mu_{\alpha}^\mathrm{*,DraII}, \mu_{\delta}^\mathrm{DraII}$] (see section 6 for more details), while the uncertainties on this vector are shown as the two shaded arrows. Transparent white dots represent bright stars ($g_0 < 17$) over the field.}
\label{density_map}
\end{center}
\end{figure}

As a sanity check, the main sequence of Dra~II can be compared to the fiducials of old and metal-poor globular clusters constructed by \citet{bernard14}. A few of those fiducials are overlaid on the CMD of Dra~II in Figure~\ref{fid}. From this figure, fiducials in the metallicity ranges $-2.4 < \FeH <-2.0$ and $-1.8 < \FeH <-1.4$  provide good visual match to the Dra~II features and its spectroscopically confirmed members (determined in section~5 below and highlighted in yellow in the figure). The most metal-poor fiducials, however, provide a better match for stars with $P_{mem} > 0.01$ brighter than $g_0 = 19$ mag. Although this does not give any precise quantitative information on the metallicity of Dra~II, it confirms the metallicity measured from the CMD-fitting  procedure and from spectroscopy.

Our spacial and CMD models can be used to estimate the Dra~II membership for each star by computing the ratio of the satellite likelihood, $\mathcal{L}_{DraII}(\vec{d})$, over the total likelihood $\mathcal{L}_{DraII}(\vec{d})  + \mathcal{L}_{MW}$. These membership probabilities are reported in Table~\ref{tbl-2} for all stars in the spectroscopic sample. The membership probability can also be used to draw the density map of the Dra~II-like stellar population. The field is binned with intervals of width 0.5 arcminutes in both X and Y. For each bin, we count the density of stars. The map is further convolved with a gaussian kernel of 2 arcminutes. To identify potential structures, the distribution of background pixels, i.e. pixels located further than 4.0$r_h$, is fitted with a gamma distribution. Pixels with a density within the upper 68, 95, and 99.85 per cent of the total background pixels distribution are represented with magenta, pink, and white contours in Figure \ref{density_map}. This map tentatively reveals the existence of an extended Dra~II-like structure over the field of view, consistent with the orientation of the major axis of the satellite. This hint of extra-tidal features could be the sign that Dra~II could be tidal disrupting. The orbit of Dra~II we infer in section 6 is consistent with the direction of these potential tails. We stress that this needs to be confirmed with a spectroscopic search for members in these regions.

\begin{figure}
\begin{center}
\centerline{\includegraphics[scale=0.45]{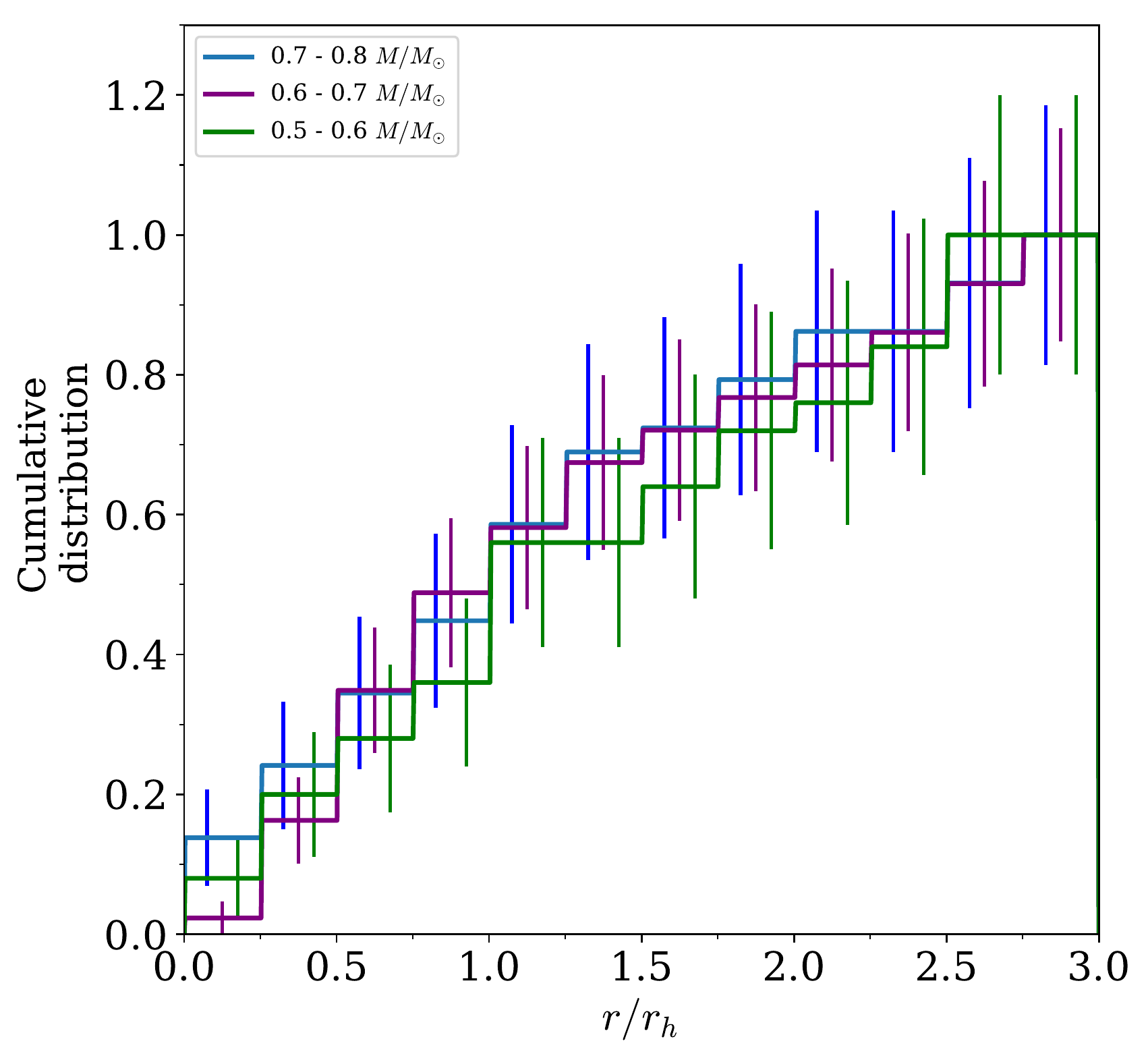}}
\caption{Normalised cumulative number of stars from 0 to 3 half-light radii, for three mass intervals : 0.8-0.7$\msun$ (blue), 0.7-0.6$\msun$ (purple) and 0.6-0.5$\msun$ (green). The analysis is performed for all stars with a CMD membership probability above 35 per cent.}
\label{mass_segregation}
\end{center}
\end{figure}

Finally, we investigate the presence of mass segregation within the system as this phenomenon can occur in globular clusters, but not in dwarf galaxies, and could therefore be used as a diagnostic for the nature of the satellite \citep{kim15}. The stellar population models provided by the Darmouth library give an estimate of the mass of a given star following these isochrones. Using this piece of information, each star within 3$r_h$ is associated with its most likely mass by comparison with the favoured isochrone.  This subsample is then separated into three mass ranges (0.5 -- 0.6$M_{\odot}$, 0.6 -- 0.7$M_{\odot}$, and 0.7 -- 0.8$M_{\odot}$). The cumulative number of stars in each mass range with respect to their radial distance to Dra~II is finally computed. This procedure is repeated for stars with a membership probability above 1, 35 and 50 per cent respectively to investigate the potential effect of the contamination on the analysis. The results are shown in Figure \ref{mass_segregation} for the 35 per cent case. This analysis gives no conclusive evidence of mass segregation in the satellite. Choosing a membership probability threshold of 1 and 50 per cent does not change significantly the results. 

\subsection{Luminosity and absolute magnitude $M_{V}$} 
We rely on the method presented in \citet{martin16_dra} to determine the total luminosity of the satellite: this method uses the PDFs on the stellar population of Dra~II and on the number of stars within the MegaCam data, $N^*$, to infer the total luminosity of the system. Therefore, it does not correspond to the sum of the fluxes of all stars seemingly members of Dra~II in the observed CMD, but it can be seen as a statistical determination of the luminosity of a system with the structural and CMD properties of Dra~II.

\begin{figure}
\begin{center}
\centerline{\includegraphics[scale=0.4]{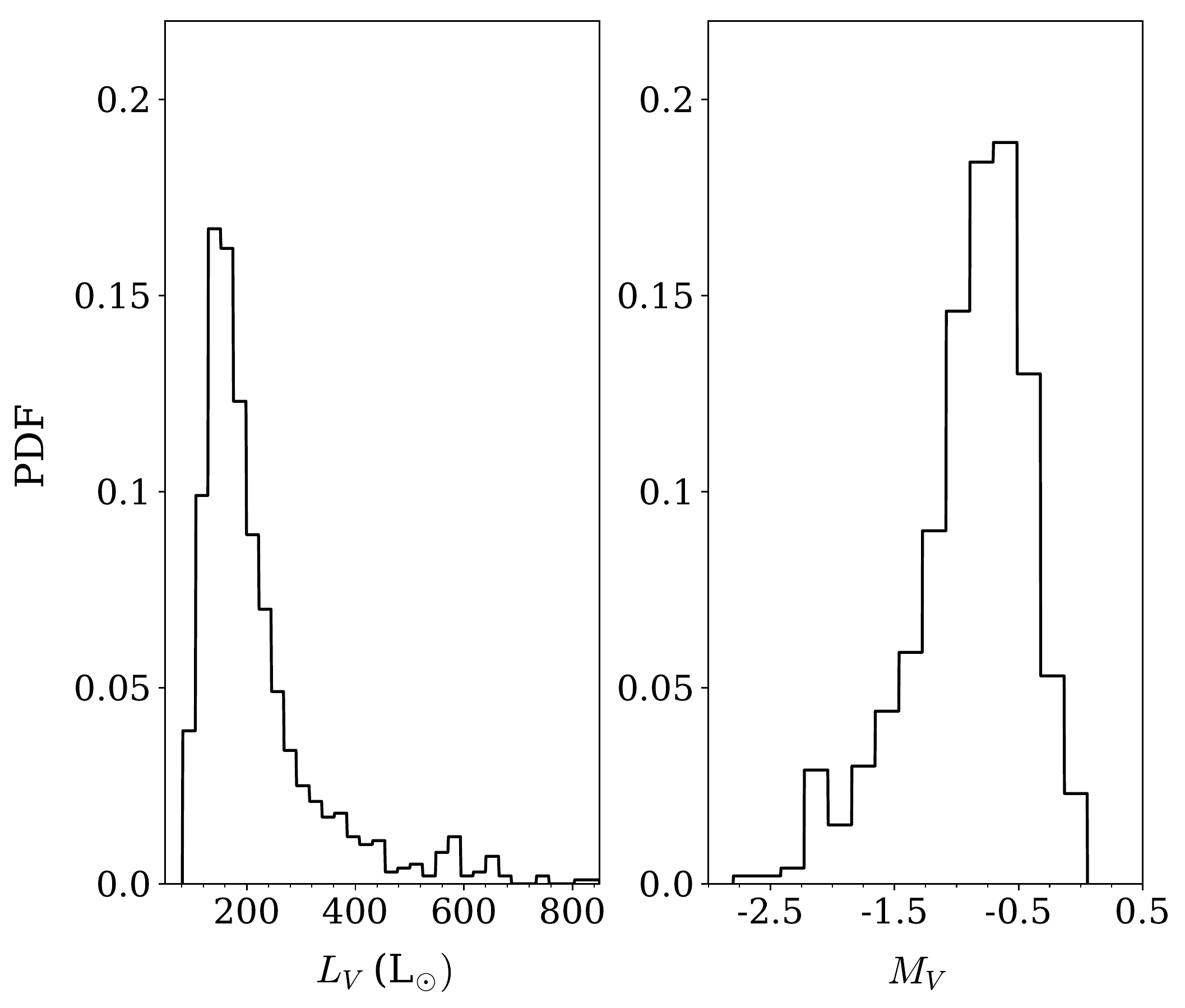}}
\caption{PDFs of the V-band luminosity (left panel) and absolute magnitude (right panel) of Dra~II. The system is particularly faint, with a favoured luminosity of only $L_V = 180 ^{+124}_{-72} L_{\odot}$.}
\label{pdfs_lum}
\end{center}
\end{figure}

At every iteration in the procedure, we randomly draw a target $N^*_j$ value from the $N^*$ PDF, as well as a set of stellar parameters ($A_j$, $[\alpha/\textrm{Fe}$]$_j$, $\FeH_{\mathrm{CMD},j}$, $(m-M)$$_j$) from the PDFs obtained through the inference of section 3.1. CMD stars are then simulated according to the j-th stellar population. The probability to draw a star at a given magnitude $g_0$ is given by the luminosity function. For each simulated star, its colour $(g-i)_0$ and magnitude $g_0$ are checked. If they fall within the CMD box used to perform the fit in the previous section, it is flagged. The simulated star is then independently checked against the completeness of the data in both $g$ and $i$. The $g$ and $i$ values are then converted into a $V$ magnitude using the colour equations presented in \citet{tonry12}. Once the number of flagged simulated stars is equal to $N^*_j$, the flux of all stars, flagged or not, is summed to yield the total luminosity, $L_{V,j}$, of that realization of the satellite. Those luminosity values are then converted into absolute magnitudes, $M_{V,j}$. Repeating this exercise several thousands of times yields the PDFs presented in Figure~\ref{pdfs_lum}.

From this analysis, Dra~II emerges as a very faint satellite, with a luminosity of only $L_V  = 180^{+124}_{-72} \; L_{\odot}$, corresponding to an absolute magnitude of $M_{V} =  -0.8^{+0.4}_{-1.0} $ mag. The satellite has a surface brightness of $\mu_0 = 28.1 \pm 0.7$ mag.arcsec$^{-2}$, comparable to the Milky-Way satellites with the lowest surface brightness. Shifting the completeness model by $\pm$ 0.5 mag does not significantly change the inferred luminosity. Most of the simulated CMDs contain no RGB star much brighter than the turnoff, which is compatible with the absence of confirmed RGB stars brighter than $g_0 = 19.3$ in our spectroscopic sample (Figure~\ref{fid} and Table 2 for the member list) and with the observed CMD. This value is however significantly fainter than the one of \citet{laevens15}, who found a luminosity of $1259^{+1903}_{-758} \lsun$. Two possible explanations for this difference can be proposed. First, the photometry at hand in 2015 is 2 magnitudes shallower than ours, thus only reaching the bright end of the main sequence of Dra~II. Finally, a small fraction of our simulated CMDs still predicts the existence of a RGB star in the satellite that could lead to a significant increase in luminosity. This is illustrated by the bright tail up to 800 L$_\odot$ in the left panel of Figure \ref{pdfs_lum} which is only due to the existence of one or two giant stars in a small fraction of our simulated CMDs. One bright star ($g < 16$) in Dra~II would potentially be enough to solve the discrepancy between \citet{laevens15} and this work. However, recent spectroscopic investigations of bright Dra~II candidates did not lead to the identification of any additional member with $g < 17$. Therefore, the discrepancy found regarding the luminosity must be caused by an overestimation of the additional overall number of stars by \citet{laevens15}, driven by shallower and noisier data.

\section{Narrow-band CaHK analysis}

The Pristine survey \citep{starkenburg17} combines CFHT narrow-band CaHK photometry with broadband colours, typically $g-i$, to infer photometric metallicities (hereafter [Fe/H]$_\mathrm{CaHK}$). A specific set of Pristine observations aims at observing all known northern Milky-Way dwarf galaxy (or dwarf-galaxy candidate) with $M_V>-9.0$. These images are much deeper than the usual Pristine observations (1-hour vs. 100-second integrations) but remain shallower than the broadband $g$ and $i$ photometry described in section~3. Reliable CaHK photometry, i.e. with $CaHK$ uncertainty below 0.1, is achieved down to $g\sim23.0$.

\begin{figure}
\begin{center}
\centerline{\includegraphics[width=\hsize]{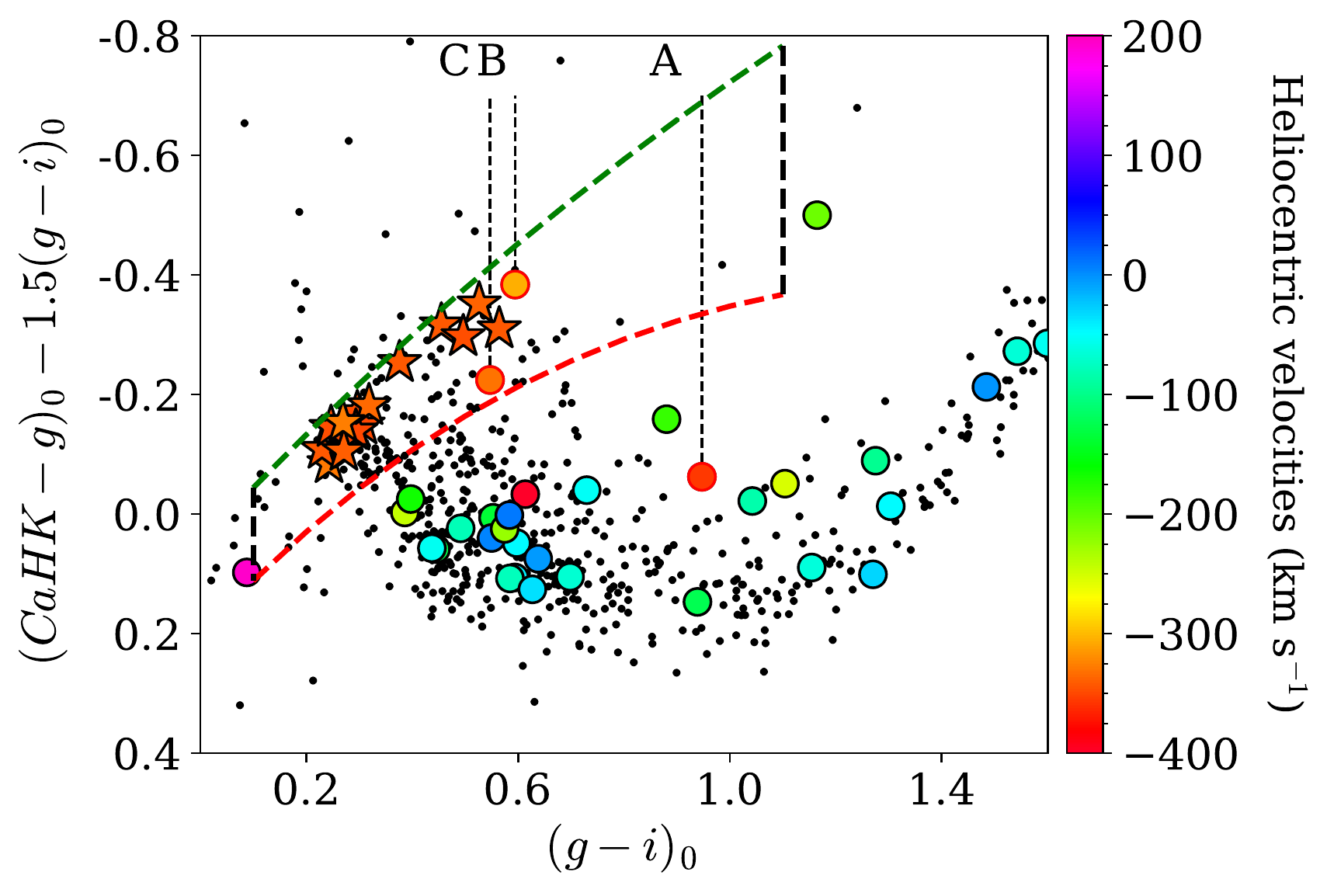}}
\caption{ Pristine colour-colour diagram. The usual temperature proxy $(g-i)_0$ is represented on the x-axis while the metallicity information is carried by the $(CaHK - g) - 1.5*(g-i)$ colour shown on the y-axis (see \citet{starkenburg17}). Stars observed spectroscopically that pass the $CaHK$ quality cut, i.e. an uncertainty on the $CaHK$ photometry below 0.1, are colour-coded according to their heliocentric velocities. Small black dots are field stars and form a clear stellar locus of more metal-rich stars ([Fe/H] $\sim -1$ or above) while more metal-poor stars are located towards above this sequence. Two iso-metallicity sequences with $\FeH = -3.5$ and $\FeH = -1.8$ are shown as green and red dashed lines, respectively. Most stars compatible with the velocity of Dra~II (red-orange) are located between these two sequences, and form a distinct, more metal-poor population than the rest of the spectroscopic sample made of more metal-rich halo and disc stars. The black dashed lines show a colour cut of $ 0.1 < (g-i)_0 <1.1$, which is applied to discard potential foreground dwarfs. A, B and C are the three stars close to the Dra~II velocity peak that were discarded using the CaHK and CMD cuts (see the text for more detail).}
\label{CaHK}
\end{center}
\end{figure}

In Figure~\ref{CaHK}, we show the typical colour-colour space used by the Pristine collaboration, for which stars with $\FeH \sim -1$ or lower reside in the bottom part of the panel and more metal-poor stars towards the top. Comparison via models and calibration onto thousands of stars in common with the Segue spectroscopic survey allow us to assign a $\FeH_\mathrm{CaHK}$ value to all these stars \citep{starkenburg17,youakim17}.  Two iso-metallicity sequences of respectively $\FeH = -3.5$ (green dashed line) and $\FeH = -1.8$ (red dashed line) are shown in the figure for illustration purposes. In the figure, we also highlight stars that are part of our DEIMOS spectroscopic sample that will be discussed in the next section. The group of likely Dra~II members at $v_r\sim-345\kms$ mainly clumps along a low metallicity sequence that is compatible with the low metallicity inferred from the broadband photometry.

The \citet{starkenburg17} Pristine metallicity model tends to slightly underestimate the metallicity at the low-metallicity end. Therefore, before turning to the Dra~II CaHK data, we first estimate and correct for this bias when determining a $\FeH_\mathrm{CaHK}$. We use the same catalogue \citet{starkenburg17} used to build their ($CaHK$, $g$, $i$) to $\FeH_\mathrm{CaHK}$ model, with the same quality criteria on the Pristine photometry and SEGUE/SDSS spectra. We bin this sample of 3,999 stars into 0.2 dex bins in metallicity for stars in the interval $-4.0 <\FeH_\mathrm{CaHK}<-1.0$. For each of these bins, we determine the median value of both $\FeH_\mathrm{SEGUE}$ and $\FeH_\mathrm{CaHK}$. The bias is then defined as the difference between these two values. This set of values is then fitted with a third-order polynomial to model the metallicity bias throughout the whole metallicity range. This bias is, at most, of $\sim0.2$\,dex at $\FeH\lta-2.0$.

\begin{figure}
\begin{center}
\centerline{\includegraphics[scale=0.55]{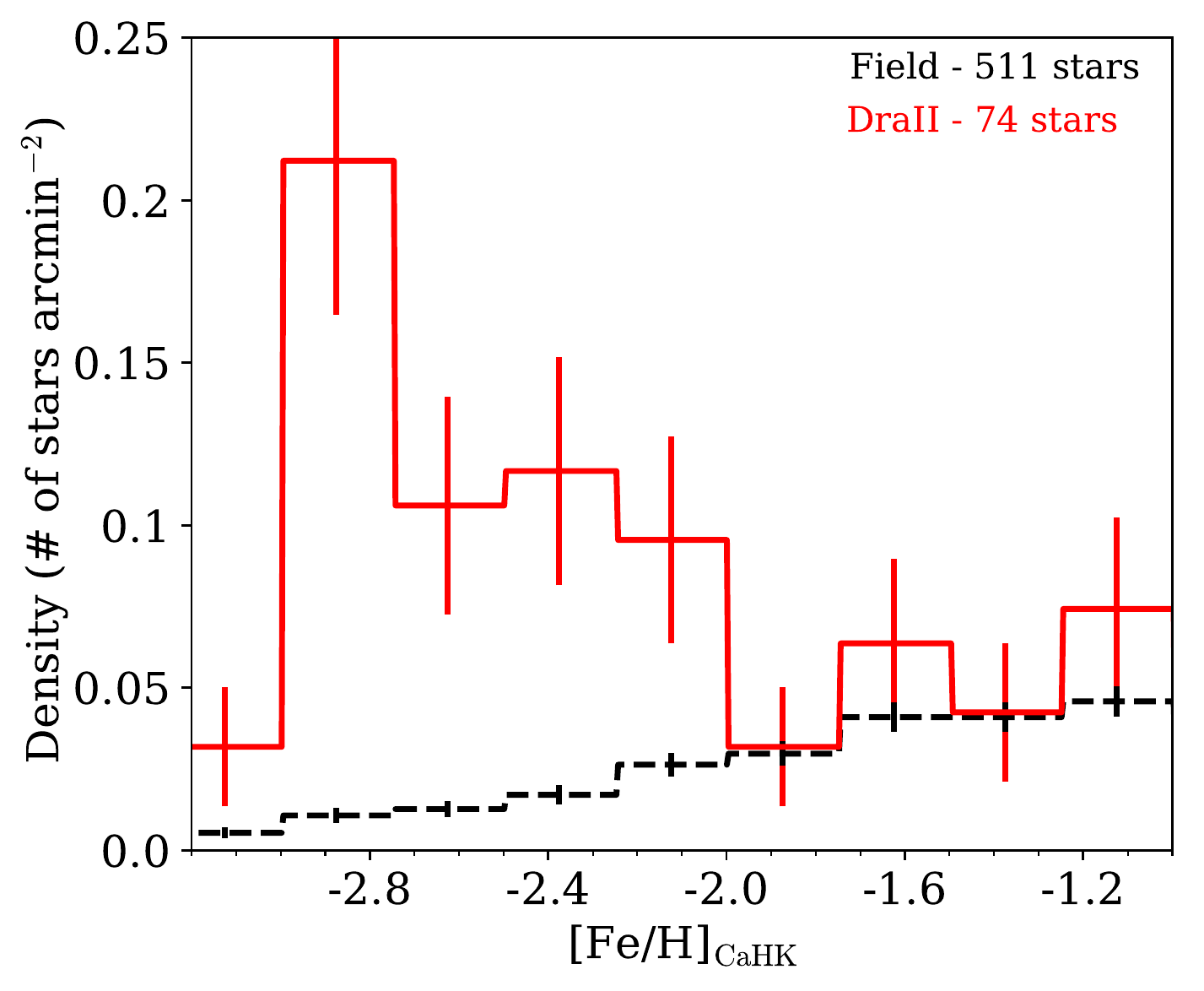}}
\caption{Area-normalised metallicity distribution for all stars within 2$r_h$ (solid red line). The same histogram is also shown for all field stars, i.e. stars outside 5$r_h$ (black dashed line). Dra~II members are clearly responsible for an overdensity of stars around $\FeH\sim -2.8$ in the red distribution.}
\label{histos_FeH}
\end{center}
\end{figure}

For every star in the Dra~II sample with uncertainties on the CaHK magnitude below 0.1, we first apply the model of \cite{starkenburg17} to infer a photometric metallicity, which we then correct for the bias modeled above. The area-normalised metallicity distribution for stars within 2$r_h$ of Dra~II is shown in red in Figure~\ref{histos_FeH}. For comparison, the black-dashed histogram shows the same distribution but for field stars, i.e. for all stars outside 5$r_{h}$. Dra~II stars stand out quite prominently as a significantly metal-poor overdensity compared to the field contamination. A lot of stars in the figure appear to be at the same photometric metallicity around -3.0. However, the calibration of the Pristine model becomes unreliable at $\FeH_\mathrm{CaHK} \sim -3.0$. The high number of stars at $\sim -3.0$ stars is probably a consequence of this.

Using Pristine metallicities, we want to infer both the mean metallicity of the system, $\langle\FeH_\mathrm{DraII}^{\mathrm{CaHK}}\rangle$, and its dispersion $\sigma_\mathrm{[Fe/H]}$. In order to do so, we assume that the distribution of photometric metallicities in the Dra~II sample corresponds to a Gaussian-distributed Dra~II population and a contamination model $\mathcal{L}_\mathrm{bkg}$, which is constructed empirically from the field data outside a 5$r_h$ radius. The metallicity distribution of this contamination sample is binned, then smoothed with a Gaussian kernel of 0.1\,dex to account for poor number counts in some metallicity bins. Finally, we assume the following metallicity distribution model:

\begin{eqnarray}
\mathcal{L}(\{\FeH_{\mathrm{CaHK},k},\delta_{\mathrm{[Fe/H]},k}\}|\langle\FeH_\mathrm{DraII}^{\mathrm{CaHK}}\rangle,\sigma_\mathrm{[Fe/H]}) \nonumber \\
= \prod_k G(\FeH_{\mathrm{CaHK},k}|\langle\FeH_\mathrm{DraII}^{\mathrm{CaHK}}\rangle,\sigma_k) + \mathcal{L}_\mathrm{bkg}(\FeH_{\mathrm{CaHK},k}),
\end{eqnarray}

\noindent with $G(x|\mu,\sigma)$ the value of a Gaussian distribution of mean $\mu$ and dispersion $\sigma$ evaluated for $x$, $\delta_{\mathrm{[Fe/H]},k}$ the uncertainty on the photometric metallicity of star $k$, and $\sigma_k = \sqrt{\sigma_\mathrm{[Fe/H]}^2+\delta_{\mathrm{[Fe/H]},k}^2}$.

\begin{figure}
\begin{center}
\centerline{\includegraphics[width=\hsize]{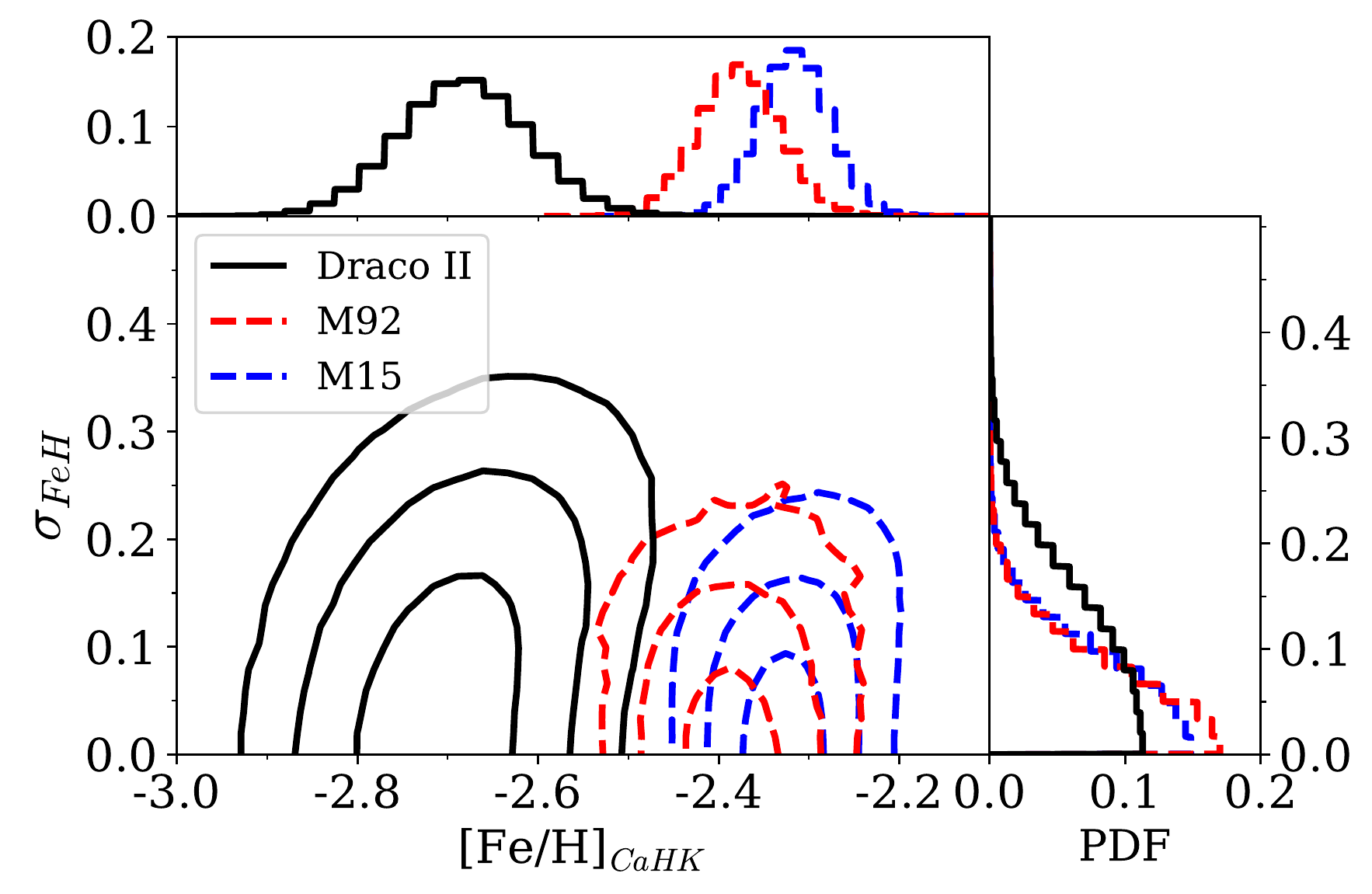}}
\caption{Two-dimensional joined PDFs of the systemic CaHK metallicity and its associated dispersion for Dra~II (black) and the globular clusters M15 and M92 (blue and red-dashed line, respectively). The marginalised one-dimensional PDFs are shown in the top and right-hand panels for the two parameters. The contours represent the usual 1, 2 and 3$\sigma$ confidence intervals in the case of a two-dimensional Gaussian distribution.}
\label{contours_FeH}
\end{center}
\end{figure}

The inference analysis yields the two-dimensional joined PDF of $\langle\FeH_\mathrm{DraII}^{\mathrm{CaHK}}\rangle$ and $\sigma_\mathrm{[Fe/H]}$ presented in Figure~\ref{contours_FeH}. The metallicity of the system is found to be $\langle\FeH_\mathrm{DraII}^{\mathrm{CaHK}}\rangle = -2.7 \pm 0.1$\,dex, with a metallicity dispersion lower than 0.24 dex at the $95$\% confidence level. The favoured systemic metallicity confirms that Dra~II is significantly metal-poor. The fraction of Dra~II stars favoured by the analysis is $\eta \sim 0.6$, corresponding to a total of 41 stars. The metallicity dispersion of the satellite cannot be resolved with this dataset. Performing the analysis using an asymmetrical gaussian does not change significantly change our results.

In order to validate our inference based on the CaHK metallicities, the same analysis is performed on the Pristine data of two metal-poor globular clusters, M15 and M92. Globular clusters are crucial in this case as their metallicity dispersion is expected to be too small to be resolved using purely photometric metallicities and they are a good test of the quality of our constraints on $\sigma_\mathrm{[Fe/H]}$. \citet[C09]{carretta09} and \citet[C09b]{carretta09b} showed that both clusters have a similar spectroscopic metallicity, with $\FeH_\mathrm{C09b} = -2.34 \pm 0.06$ dex for M15 and $\FeH_\mathrm{C09} = -2.35 \pm 0.05$ dex for M92, as well as metallicity dispersions around $\sim 0.05$. The application of our inference model to the globular cluster Pristine datasets yields $\langle\FeH_\mathrm{M15}\rangle=-2.32\pm0.04$ dex and $\langle\FeH_\mathrm{M92}\rangle =-2.38 \pm 0.05$ dex, compatible with the values of C09. As expected, the inferred metallicity dispersions are unresolved for both clusters, as can be seen with the coloured contours in Figure~\ref{contours_FeH}. The favoured models yields 43 stars for M15 and 25 stars for M92, comparable to the 41 stars studied in Dra~II.

The inference on the metallicity mean and dispersion for the two globular clusters is as expected and yields confidence that the CaHK metallicities are reliable. We therefore conclude that Dra~II is indeed a very metal-poor satellite and we further note that despite similar numbers of member stars in the three systems, the Dra~II metallicity dispersion PDF is wider than that of the clusters, which may hint at a larger metallicity dispersion for Dra~II.

\section{Spectroscopic analysis} 
\label{spectro_sec}
\begin{figure*}
\begin{center}
\centerline{\includegraphics[width=\hsize]{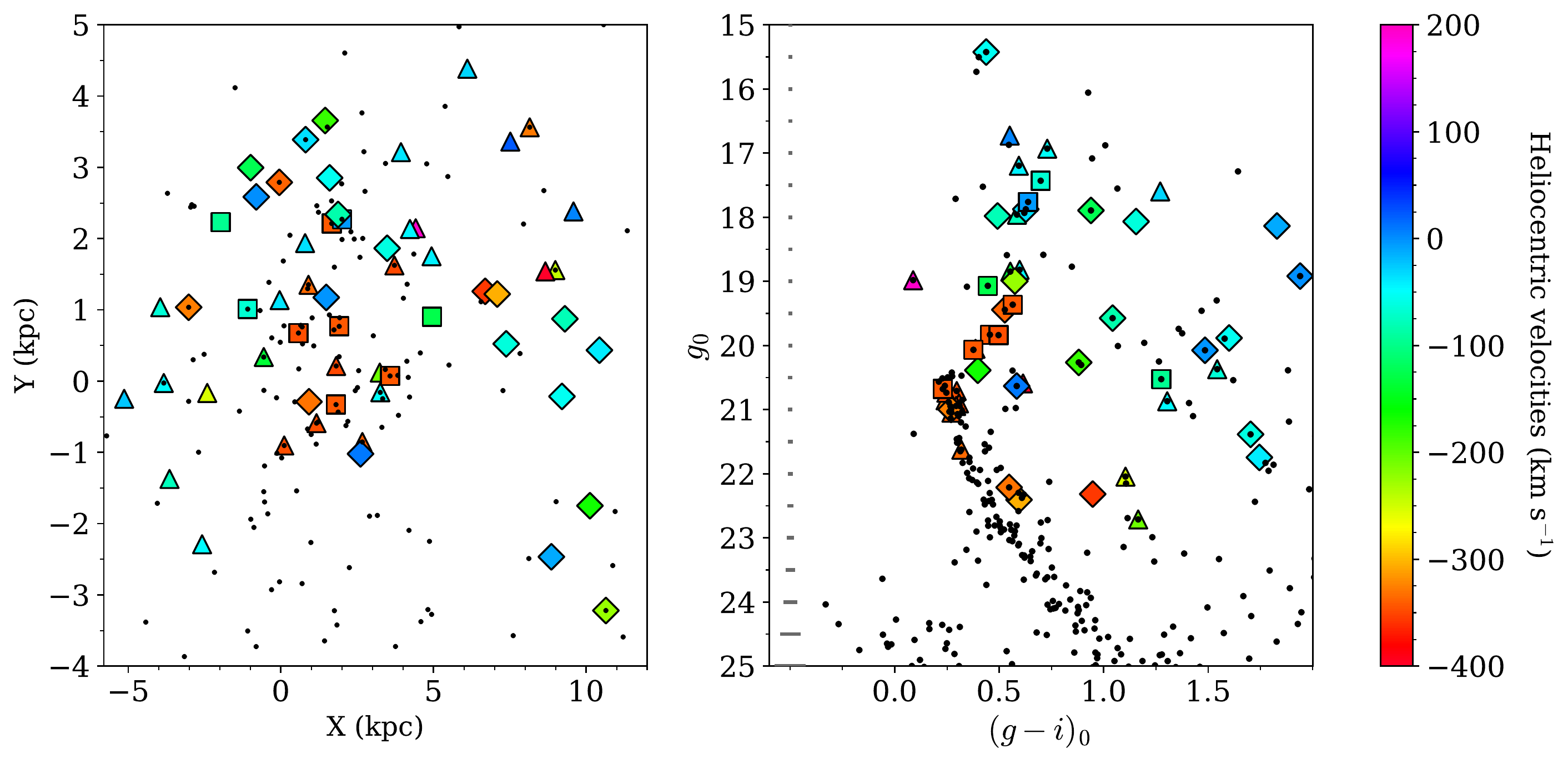}}
\caption{\textit{Left panel:} Magnified view of the central region showing the spacial distribution of the spectroscopic sample. Stars observed spectroscopically are represented by large dots colour-coded according to their heliocentric velocities. Diamonds and triangles correspond to stars observed in 2015 and 2016 respectively, while squares correspond to stars observed both in 2015 and 2016. Some of the stars observed spectroscopically do not overlap small black dots as those correspond only to Dra~II like population and do not represent the full photometric dataset. \textit{Right panel:} Distribution of the spectroscopic sample in the CMD within two half-light radii of Dra~II. Stars lying on the Dra II main sequence, in red-orange, are likely members of the system. Some of the stars with velocity measurements are located further away than 2 half-light radii and thus do not also appear as small dots.}
\label{spectro_plot}
\end{center}
\end{figure*}

\begin{figure}
\begin{center}
\centerline{\includegraphics[width=\hsize]{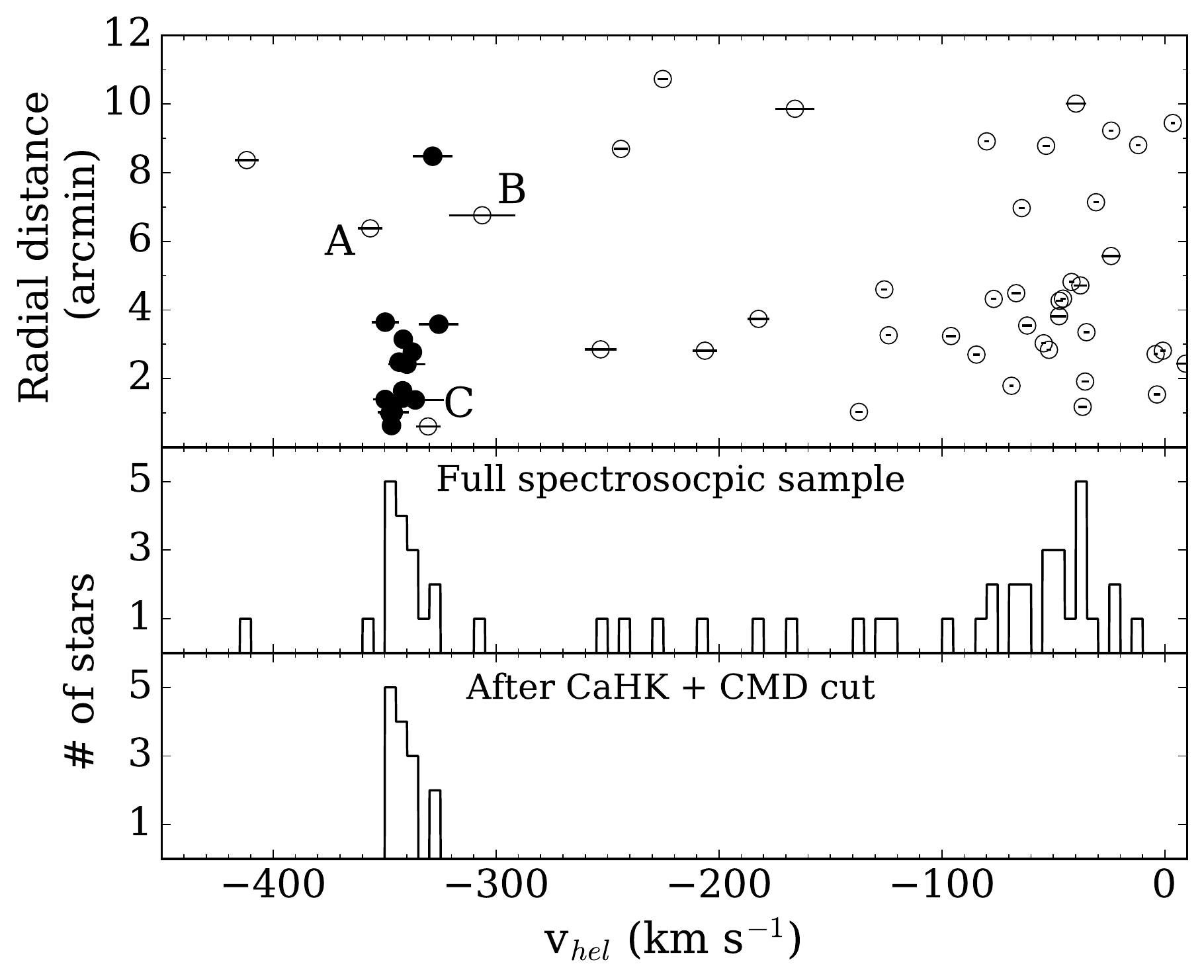}}
\caption{\textit{ Top panel:} Radial distances to the centroid of Dra~II versus heliocentric velocities for all stars in our spectrosopic sample. Black-filled markers represent the remaining spectroscopic population after the CaHK and CMD cuts were applied. They are considered as dynamical members of the system. \textit{Middle panel:} Histogram of the heliocentric velocities in the spectroscopic sample. \textit{Bottom panel:} Histogram of velocities for dynamical member stars only, obtained by discarding stars that do not come out as metal-poor through the CaHK model detailed in section 4 (Figure~\ref{CaHK}), as well as stars that are not compatible with our favoured CMD model (section 3). They correspond to the black-filled markers in the top panel.}
\label{vel_histo}
\end{center}
\end{figure}

We now investigate the dynamical properties of the satellite using our spectroscopic data, for which the processing was detailed in section 2.2. Examples of spectra can be found in Figure 4 of \citet{martin16_dra}, who display 4 spectra of our 2015 run that are representative of the whole dataset since the 2016 spectroscopic observations were performed under similar conditions and have similar quality. Their spacial and CMD distribution are shown in Figure \ref{spectro_plot}. The histograms of heliocentric velocities for our 2015 and 2016 runs combined are shown in the middle panel of Figure~\ref{vel_histo}.

Dra~II stars clearly stand out in Figure~\ref{vel_histo} as they form a peak around $-345\kms$, as was already pointed out by \citet{martin16_dra} in their initial analysis of the 2015 data set. A broader distribution around $\sim-45\kms$ corresponds to stars from the Milky Way disc while Milky Way halo stars are responsible for the sparsely distributed velocities throughout the range shown here. In order to better constrain the dynamical properties of the system, one has to isolate Dra~II members as well as possible. Particular care should be taken when handling the contamination by Milky Way halo stars that are distributed within a broad velocity range that includes the systemic velocity of the satellite. For this reason, it would not be surprising to find a few contaminating stars in the vicinity of the velocity peak of Dra~II. In particular, one can also notice the existence of two slight outliers around the Dra~II velocity peak, noted stars A and B. It is quite challenging to know whether those stars are bona fide members based only on their kinematic properties. This is a common problem when dealing with such faint systems for which only a handful of members are confirmed: the velocity dispersion and systemic velocity can be biased by slight outliers that are in fact not members \citep{mcconnachie_cote10}.

Pristine CaHK photometry can be very useful to clean the spectroscopic sample as the Dra~II stellar population is very metal-poor, as shown in section 3.1. All Dra~II members are too faint to yield reliable spectroscopic metallicities but it is expected that they can be disentangled from Milky-Way contaminants by using the CaHK photometric metallicities described in the previous section.

The Pristine colour-colour diagram presented in Figure~\ref{CaHK} highlights the location of stars with heliocentric velocities, which are colour-coded according to those. As mentioned before, the metal-rich stars from the disc form a clear stellar locus at the bottom of the panel, whereas metal-poor stars are always located above this locus. The two iso-metallicity sequences that bracket the metallicity peak visible in Figure~\ref{histos_FeH}, with $\FeH=-3.5$ and $\FeH=-1.8$, are represented by the green- and red-dashed line, respectively. As expected, stars with velocities compatible with the systemic velocity of Dra~II (red-orange) are clearly isolated from the metal-rich, foreground contamination from the Milky Way. Most of the other stars from the spectroscopic sample lie in or close to the metal-rich stellar locus. To help discriminate between Dra~II stars and the contamination in our spectroscopic sample, we isolate stars with good CaHK photometry ($\delta_\mathrm{CaHK}<0.1$), metallicity uncertainties below 0.3\,dex, and, following Figure~\ref{histos_FeH}, with $-3.5<\FeH_\mathrm{CaHK}<-1.8$. Further applying a CMD-cut along the favoured isochrone of section~3.1 yields the cleaned velocity sample that is presented in the bottom panel of Figure~\ref{vel_histo}. It is obvious that the combined CaHK and CMD information has significantly cleaned the velocity distribution, leaving only highly probable Dra~II stars. As a result, star A is clearly not a member: not only is it far from the Dra~II sequence in the CMD, but it is also far too metal-rich to belong to the system. Star B seems to be at the appropriate photometric metallicity  to be a Dra~II member but is offset from the Dra~II main sequence by 0.1 mag in the CMD. This location corresponds to a part of the CMD where one might expect to find Dra~II binary stars \citep{romani91}, which could mean that this star is a Dra~II member in a binary star and therefore not reliable for the velocity analysis. We also conservatively discard star C for the same reason, even though it falls within the Dra~II velocity peak. Keeping star C or discarding it does not change our results on the velocity properties of Dra~II. Including B in the sample also has no significant impact as its velocity uncertainty is large ($\sim 15 \kms$). 

In order to derive the systemic velocity and velocity dispersion of Dra~II from this clean sample, we follow the framework of \citet{martin18} and assume stars are normally distributed. The likelihood function is therefore

\begin{equation}
\mathcal{L}(\{v_{\mathrm{r},k},\delta_{\mathrm{v},k}\}|\langle v_\mathrm{r} \rangle,\sigma_\mathrm{v})= \prod_k G(v_{\mathrm{r},k} | \langle v_\mathrm{r} \rangle,\sigma_\mathrm{k}),
\end{equation}

\noindent with $G(x|\mu,\sigma)$ the value of a Gaussian distribution of mean $\mu$ and dispersion $\sigma$ evaluated on $x$, $\delta_{\mathrm{v},k}$ the uncertainty on the photometric metallicity of star $k$ and $\sigma_k = \sqrt{\sigma_\mathrm{v}^2 + \delta_{\mathrm{v},k}^2 + \delta_{\mathrm{v},sys}^2 } $, $\delta_{\mathrm{v},sys}  $ is the systematic uncertainty floor tied to DEIMOS observations. Here we use the value determined by \citet[$\delta_{\mathrm{v},sys}  = 2.3 $ km s$^{-1}$]{martin16_dra}, which is compatible with the value we determine from the few stars in common between the Dra~II 2015 and 2016 samples.

\begin{figure}
\begin{center}
\centerline{\includegraphics[scale=0.5]{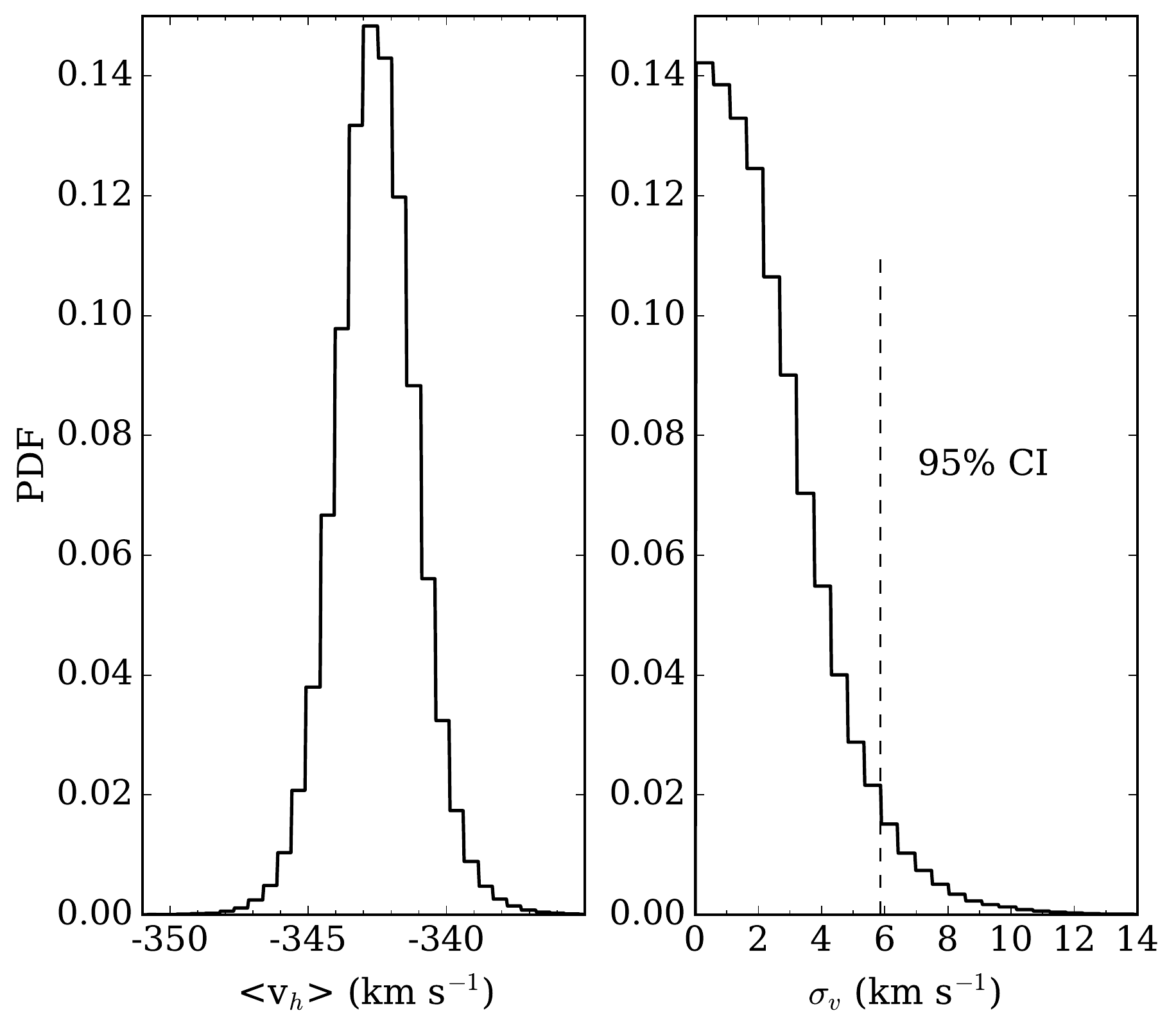}}
\caption{Marginalised PDFs for the Dra~II systemic velocity (left panel) and velocity dispersion (right panel). The system is dynamically cold, with a marginally resolved velocity dispersion. We constrain the velocity dispersion to be lower than 5.9 $\kms$ at the 95 per cent confidence level (dashed vertical line).}
\label{vel_pdfs}
\end{center}
\end{figure}

The resulting 1D PDFs of the velocity dispersion and systemic velocity are shown in Figure~\ref{vel_pdfs}. These updated results do not change significantly from those presented by \citet{martin16_dra}, despite our slightly larger  sample and the removal of dubious members by using the CaHK photometric metallicities. The velocity dispersion of Dra~II is only marginally resolved, whereas the inferred systemic velocity is $\langle v_{r}\rangle  = -342.5^{+1.1}_{-1.2}\kms$. Assuming a mass-to-light ratio (M/L) of 2 typical of MW globular clusters \citep{mclaughlin05}, a Dra~II-like GC with a size of $\sim 19$ pc and absolute magnitude of $\sim -0.8$ mag is expected to have a velocity dispersion of the order of $\sim 0.25 \kms$ if it is in equilibrium and unaffected by binaries, using the relation of \citet{walker09}. Therefore, even with a dispersion as small as $\sim 1 \kms$, Dra~II would still possess a significant amount of DM but, unfortunately, the radial velocities of the 14 members do not constrain the M/L ratio of the satellite.

\section{Gaia DR2 proper motions and orbit}

\begin{figure}
\begin{center}
\centerline{\includegraphics[scale=0.5]{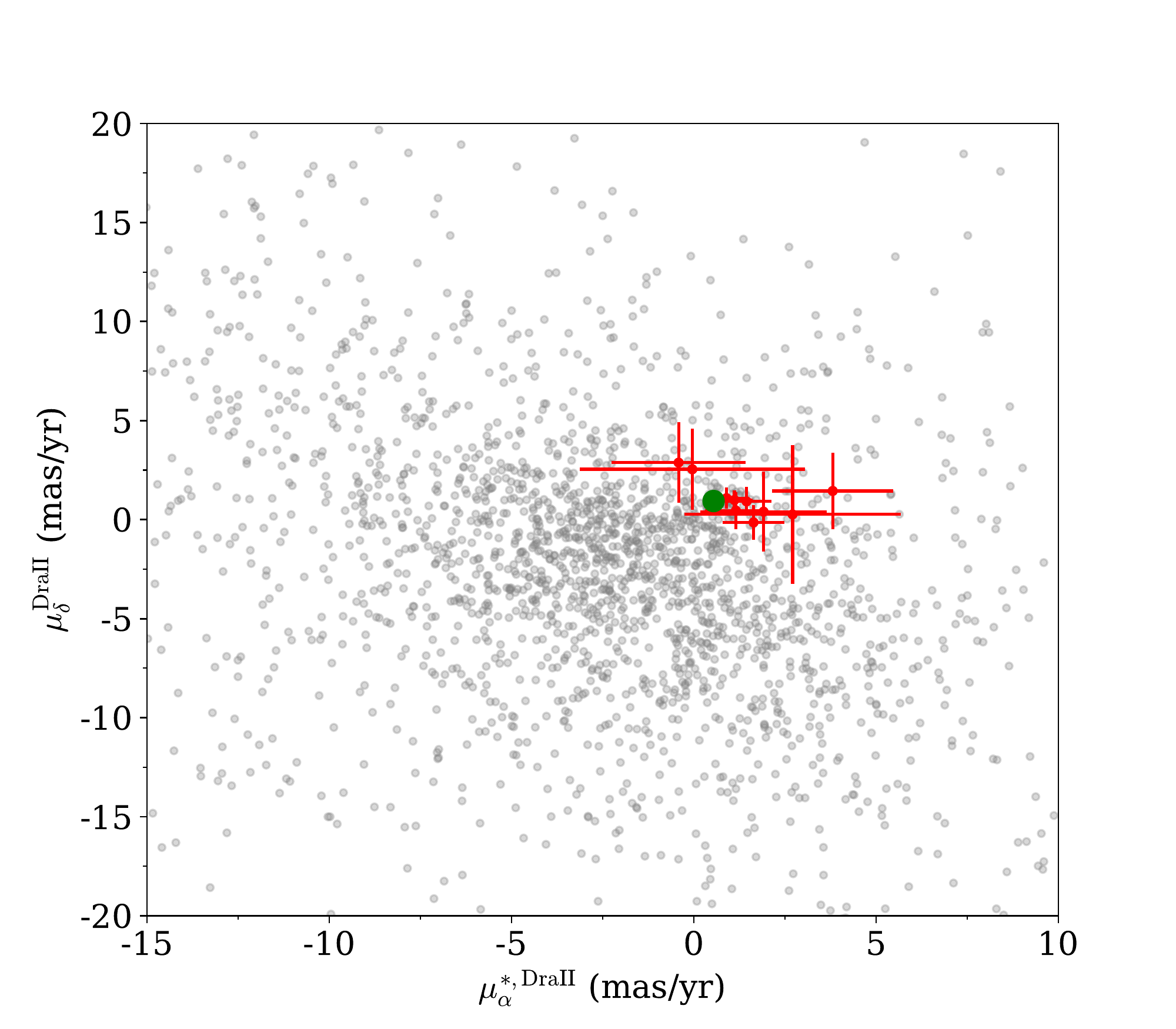}}
\caption{PMs in RA and DEC, for field stars (grey) and 10 Dra~II dynamical members (red). The mean proper motion of the satellite is represented as a large green dot.}
\label{proper_motions}
\end{center}
\end{figure}

To determine the orbit of Dra~II, we extract the proper motions (PMs) of all stars within half a degree from Dra~II's centroid in the  Gaia Data Release 2 \citep{brown18}. A cross-match between the 14 identified member stars in section 5 is then performed, resulting in 10 members with a PM measurement. The PMs of the 10 Dra~II members are shown in red in Figure \ref{proper_motions}. The uncertainty-weighted PM of Dra~II yields $\mu_{\alpha}^\mathrm{*,DraII} = \mu_{\alpha}^\mathrm{DraII} \cos(\delta) = 1.26 \pm 0.27$ mas.yr$^{-1}$ and $\mu_{\delta}^\mathrm{DraII} = 0.94 \pm 0.28$ mas.yr$^{-1}$, and is shown in Figure \ref{proper_motions} as the large, green dot. These measurements take into account the systematic error of 0.035 mas.yr$^{-1}$  on the PMs for dSph as shown by \citet{helmi18}. However, if we instead choose the systematic error presented in that paper for the GCs, our results do not change given the measured uncertainties on $\mu_{\alpha}^\mathrm{*,DraII}$ and $\mu_{\delta}^\mathrm{DraII}$.

\begin{figure*}
\begin{center}
\centerline{\includegraphics[scale=0.5]{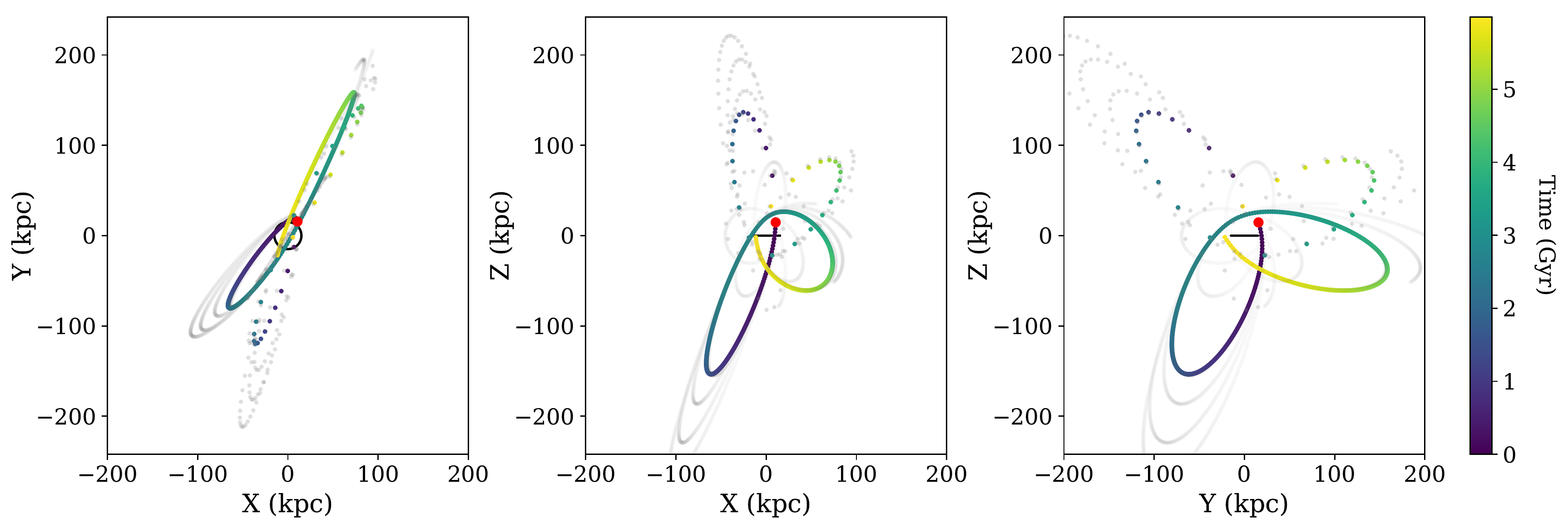}}
\caption{Projections of the orbit of Dra~II on the X-Y, X-Z and Y-Z planes backwards and forwards over 6.0 Gyr. Six orbits are shown here: the one based on the favoured position, distance, radial velocity and PMs of the satellite, and five others using random realizations of those parameters, as slightly transparent, grey lines. The median orbit is colour-coded according to the time elapsed since present day, in Gyr. Dotted lines represent the backwards-integrated orbits. The current position of Dra~II is indicated with a red dot, while the MW disk is shown in black, with a chosen radius of 15 kpc.}
\label{orbits}
\end{center}
\end{figure*}

These measurements can be used to put constraints on the orbit of the satellite. To do so, we rely on the GALPY package \citep{bovy15}. The MW potential chosen to integrate Dra~II orbit is based on the so-called ``MWPotential14'' defined within GALPY, constituted of three components: a power-law, exponentially cut-off bulge, a Miyamoto-Nagai potential disk, and a Navarro-Frenk-White DM halo. A more massive halo is chosen for this analysis, with a mass of $1.2 \times 10^{12} \msun$ (vs. $0.8 \times 10^{12} \msun$ for the halo used in MWPotential14).  We integrate 1000 orbits backwards and forwards over 6 Gyr, each time by randomly drawing a position, distance, radial velocity, and PMs from their corresponding PDFs, and extract for each realization the apocenter, pericenter, and ellipticity. The orbit of the favoured model (i.e. favoured position, distance, radial velocity and PMs) is shown in Figure \ref{orbits} in the X-Y, X-Z and Y-Z planes, and colour-coded by time. Five random realizations of the orbit are also shown in this figure as partially transparent, grey lines.  

This analysis yields a pericenter of $21.3 ^{+0.7}_{-1.0}$ kpc, an apocenter of $153.8 ^{+56.7}_{-34.7}$ kpc and an ellipticity of $0.77 ^{+0.08}_{-0.06}$. Dra~II seems to be on a quasi-perpendicular orbit with respect to the disk of the MW. Our orbit is compatible with the one of \citet{simon18}, though they favour a slightly higher apocenter due to their use of a light MW ($0.8 \times 10^{12} \msun$). This is confirmed by the analysis of \citet{fritz18}, whose results are also consistent with ours. Our larger sample nevertheless provides a more stringent constraint on the orbit of Dra~II. The fairly elliptical orbit and the small pericenter we infer appear compatible with the idea that the satellite has been severely affected by tides and could explain the low-surface brightness features seen in Figure~\ref{density_map} that roughly align with the PM vector. 

\section{Summary and discussion}
In this paper, we present an analysis of our deep MegaCam/CFHT broad-band $g$ and $i$ photometry of Dra~II, combined with narrow-band CaHK photometry from a specific sub-program of the Pristine survey that focuses on all northern sky dwarf-galaxy candidates. We also present an analysis of the extension of our multi-object spectroscopy observed with Keck~II/DEIMOS.

We estimate the structural parameters of Dra~II and infer properties that are compatible with the previous study of the satellite by \citet{laevens15} albeit with smaller uncertainties: the system has a half-light radius of $r_h = 19.0 ^{+4.5}_{-2.6}$ pc and is remarkably faint ($L_V = 180 ^{+124}_{-72} L_{\odot}$). Based on the CMD information of the observed stars, we confirm that Dra~II hosts an old stellar population with an age of $13.5 \pm 0.5 \Gyr$, a metallicity $\FeH_\mathrm{CMD} = -2.40 \pm 0.05$ dex, $[\alpha/\textrm{Fe}] = +0.6$ dex, and a distance modulus of $m-M=16.67\pm 0.05$ mag. Using the Pristine photometry, we were able to find an estimate of the metallicity of Dra~II with $\langle\FeH_\mathrm{DraII}^{\mathrm{CaHK}}\rangle=-2.7\pm 0.1$ dex. This inference is confirmed by the analysis of 3 Dra~II spectroscopic members, which yields $\FeH_{\mathrm{spectro}} = -2.43 ^{+0.41}_{-0.82}$ dex. The metallicity derived from the three different techniques are therefore all consistent. However, the isochrone fitting procedure is limited by the model grid, for which the lowest metallicity is $-2.45$ dex. Three low-RGB stars were used to derive the spectroscopic metallicity of the satellite using the Calcium triplet relation of \citet{starkenburg10}. However, this relation is calibrated for RGB stars, though \citet{leaman13} shows that it can give consistent results when applied to stars 2 magnitudes fainter. We therefore favour the systemic metallicity inferred by the CaHK technique as it does not suffer from these limitations. The metallicity dispersion of Dra~II is only marginally resolved for both the spectroscopic and CaHK procedures. Similarly, applying the same technique to the two old and metal-poor globular clusters M15 and M92 yields no measurable metallicity dispersion, in line with expectations for globular clusters. Finally, we combined the CaHK and broad-band information with our DEIMOS spectroscopy to isolate 14 likely member stars. This sample is used to derive a systemic velocity of $\langle v_{r}\rangle = -342.5^{+1.1}_{-1.2}\kms$ and a marginally resolved velocity dispersion, confirming that Dra~II is a particularly cold system. Finally, using the Data Release 2 of Gaia, we use 10 Dra~II member stars to characterize the orbit of the system: the apocenter and pericenter are found to be $153.8 ^{+56.7}_{-34.7}$ kpc and $21.3 ^{+0.7}_{-1.0}$ kpc respectively.

\begin{figure*}
\begin{center}
\centerline{\includegraphics[width=\hsize]{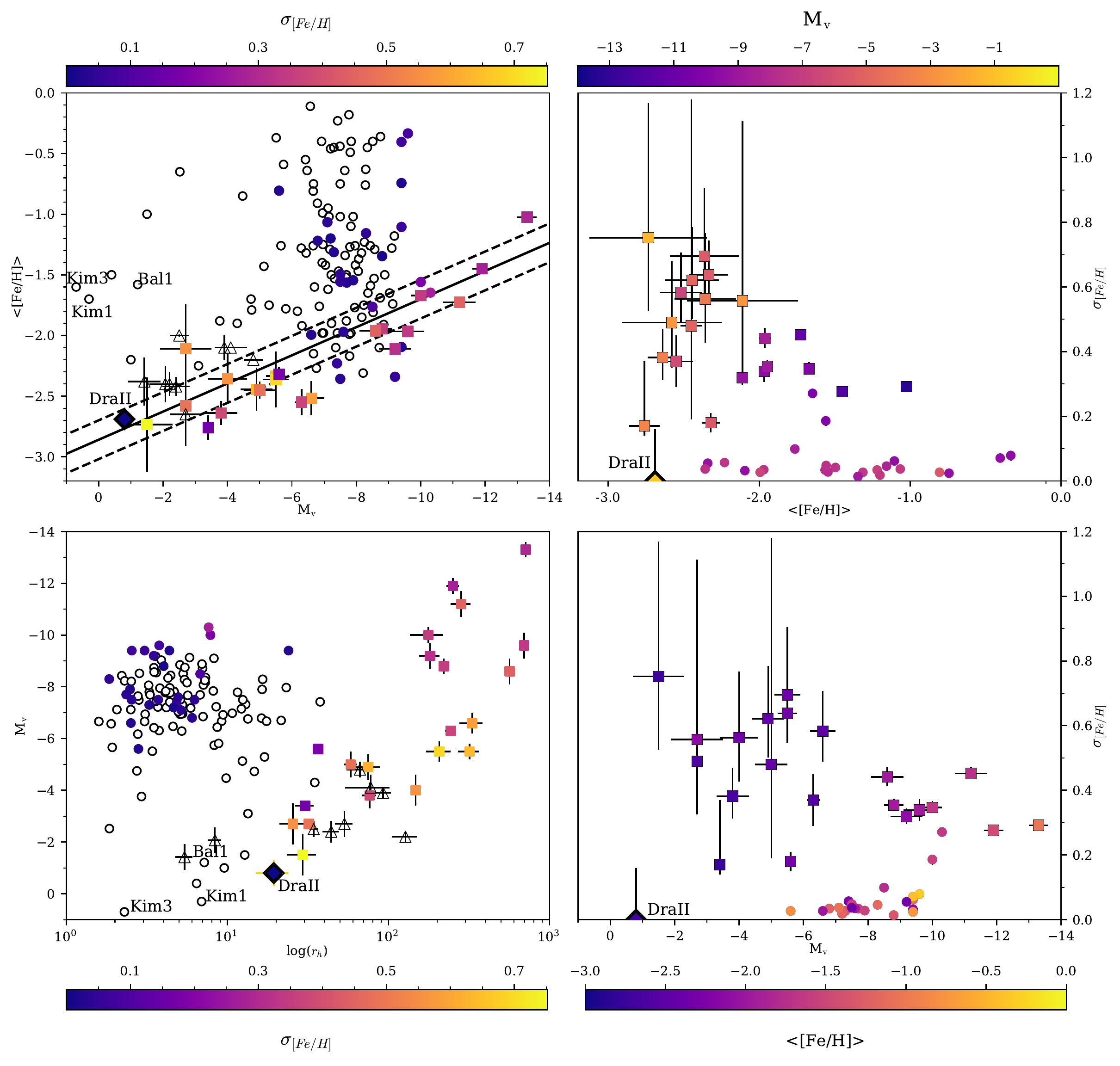}}
\caption{ Comparison of Dra~II with other GCs and dwarf galaxies of the Milky Way. Squares represent dwarf galaxies while circles represent globular clusters, and the diamond corresponds to Dra~II. Triangles stand for recently discovered dwarf-galaxy candidates that await confirmation. Hollow markers correspond to systems for which no metallicity dispersion measurement can be found in the literature. The solid line in the top-left panel corresponds to the luminosity-metallicity relation of \citet{kirby13} for dwarf spheroidals and dwarf irregulars. Dashed lines represent the RMS about this relation, also taken from \citet{kirby13}. Among the 123 globular clusters presented here, the properties of 116  were extracted from \citet{harris96} catalog, revised in 2010. For the remaining ones (Kim 1, Kim 2, Kim 3, Laevens 1, Balbinot 1, Munoz 1 and SMASH 1) parameters of the discovery publications were used (\citet{kim15b}, \citet{kim15}, \citet{kim16b}, \citet{laevens14}, \citet{balbinot13}, \citet{munoz12b} and \citet{martin16b}). Globular cluster metallicity spread measurements are taken from \citet{willman_strader12} and references therein: \citet{carretta06,carretta07b,carretta09b,carretta11}, \citet{cohen10}, \citet{gratton07}, \citet{johnson_pilachowski10}, and \citet{marino11}. \citet{mcconnachie12} and \citet{willman_strader12} are used to compile the properties of the dwarf galaxies represented here. The 18 dwarf galaxies represented here are: Bootes I \citep{belokurov06,norris10}, Canes Venatici I \citep{zucker06b}, Canes Venatici II \citep{sakamoto06}, Coma Berinices, Hercules, Leo IV and Segue I \citep{belokurov07}, Draco and Ursa Minor \citep{wilson55}, Fornax \citep{shapley38b}, Leo I and Leo II \citep{harrington_wilson50}, Pisces II \citep{belokurov10}, Sculptor \citep{shapley38a}, Sextans \citep{irwin90}, Ursa Major I \citep{willman05b}, Ursa Major II \citep{zucker06a}, Willman I \citep{willman05a}. Their metallicity and metallicity spreads were drawn from \citet{kirby08}, \citet{kirby10}, \citet{norris10}, \citet{willman11}. The dwarf galaxy candidates discovered recently and shown on this figure are Bootes II \citep{koch_rich14}, DES1 \citep{luque16,conn18}, Eridanus III \citep{bechtol15,conn18,koposov15}, Hyades II \citep{martin15}, Pegasus III \citep{kim15b}, Reticulum II and Horologium I \citep{koposov15b}, Segue II \citep{belokurov09}, and the most significant candidates of \citet{drlica-wagner15} : Gru II, Tuc III, and Tuc IV.}
\label{context}
\end{center}
\end{figure*}

Despite the deep photometry studied here and the additional spectroscopy, the derived properties of Dra~II are still challenging to interpret and the nature of the system remains uncertain. Dra~II is placed in the general context of Milky Way satellites in Figure~\ref{context} and, below, we discuss two broad scenarios: whether Dra~II is a globular cluster or a dwarf galaxy.

\subsection{Is Dra~II a globular cluster?}
Figure \ref{context} (top-right panel) shows Dra~II does not present any clearly constrained dispersion in metallicity, in contrast to confirmed dwarf galaxies. Similarly, dwarf galaxies tend to be dynamically hot whereas the spectroscopic analysis of 14 Dra~II members only yields a marginally resolved velocity dispersion. These two properties are compatible with the globular cluster hypothesis. 

The globular cluster scenario does not come without difficulties, though. In particular, if the system contains no dark matter, its potential well is entirely determined by its very few stars. Using the formalism of \citet{innanen83}, the instantaneous tidal radius $r_t$ of a Milky Way satellite of mass $M_\mathrm{cluster}$ at a distance of $R$, is given by

\begin{equation}
r_\mathrm{t} = 0.43(\frac{M_\mathrm{cluster}}{M_\mathrm{MW}})^{1/3} R,
\end{equation}

\noindent with $M_\mathrm{MW}$ the mass of the Milky Way enclosed within that radius $R$.

Using Dra~II's galactocentric distance ($R \sim23.5\kpc$) and a cluster mass of 360$M_{\odot}$, obtained from the measured luminosity of Dra~II and assuming a mass-to-light ratio of 2 \citep{bell_dejong01}, the tidal radius of Dra~II is then a mere $\sim10\pc$, i.e. much smaller than the extent of a system with a measured half-light radius of $19.0 ^{+4.5}_{-2.6}\pc$. We would then be observing Dra~II just as it is being destroyed by the Milky Way's tides and, likely, on its final passage around the Galaxy. It would mean that we are observing Dra~II during a unique and short-lived moment of its lifetime, and would be a way to explain the relatively high size of the satellite compared to globular clusters of similar faintness : Kim~1 ($r_h \sim 7$ pc), Kim~3 ($r_h \sim 2$ pc) and Bal~1 ($r_h \sim 7$ pc) discovered recently \citep{conn18,koposov07,luque16,martin16b}.

Finally, the absence of any sign of mass segregation, which could occur in self-gravity dominated systems such a globular clusters \citep{kim15}, could also cast doubt on the globular-cluster nature of Dra~II, even though the existence of mass segregation, especially in a GC possibly in the midst of disruption, is not certain.

\subsection{Is Dra~II one of the faintest dwarf galaxies?}
The top-left panel of Figure~\ref{context} showcases that Milky Way dwarf galaxies follow a reasonably well-defined luminosity-metallicity relation (see also, e.g., \citealt{kirby13}). For an extremely faint stellar system like Dra~II, one would expect its metallicity to be very low ($\FeH\sim-2.5$) if it were a dwarf galaxy, which is compatible with our results using independently the CaHK photometry ($\FeH_\mathrm{CaHK} = -2.7 \pm 0.1$ dex) and spectroscopy of 3 low-RGB member stars ($\FeH_\mathrm{spectro} =  -2.43 ^{+0.41}_{-0.82}$ dex). Given the scatter and possible stochastic effects of the metallicity-luminosity relation of dwarf galaxies \citep{revaz18}, Dra~II is entirely compatible with this relation. Moreover, Dra~II has a size larger than the vast majority of known Milky Way globular clusters and, in particular, it is several times more extended than GCs of roughly the same luminosity and metallicity, as mentioned in the last section (bottom-left panel of Figure \ref{context}).

The inference on the metallicity dispersion of the satellite was performed through the spectroscopic analysis of 3 low-RGB member stars, and a new technique using the photometric CaHK metallicities of 41 stars. Though both methods do not resolve a significant metallicity dispersion, the final results do not rule it out for the system, because of the faintness and low number of the stars used in both analyses. Similarly, the velocity dispersion PDF of Dra~II does not rule out a dynamical mass higher than expected from a globular cluster: if we use the \citet{walker09} equation 4 to estimate the dynamical mass, assuming a mass-to-light ratio of 2 that is typical for globular clusters \citep{bell_dejong01}, and the luminosity and half-light radius inferred in this work, a Dra~II-like globular cluster should have a velocity dispersion around $\sim 0.25 \kms$. It is therefore challenging to determine whether the satellite has a higher dynamical mass than traced by its stars alone, as even a velocity dispersion of the order of $1 \kms$ would indicate that Dra~II has a DM halo.

Figure \ref{density_map} highlights that there might be extended tidal structures around Dra~II, consistent with the proper motion vector and the major-axis of the satellite. Its metallicity is still compatible with brighter dwarf galaxies following the metallicity-luminosity relation. Moreover, \citet{penarrubia08} showed that the velocity dispersion of a disrupted dwarf galaxy tends be lower than the original progenitor, consistent with the fact that Dra~II appears to be dynamically cold. Could Dra~II then be the final remnant of a  brighter dwarf galaxy that lost 90 per cent of its mass through tidal interactions with the Milky Way? Such a disruption would not be surprising given the orbit of the satellite, with a pericenter of $21.3 ^{+0.7}_{-1.0}$ kpc.

\subsection{Conclusion}
The properties of Draco~II tend to indicate that the satellite is a potentially disrupting dwarf galaxy, which could explain its total luminosity, metallicity, size, and relatively low velocity dispersion. The orbit of the satellite, constrained with Gaia PMs, shows that the satellite is very likely to be affected by tidal processes, which is backed up by potential tidal structures observed in the field. However, the impossibility, with our current dataset, to constrain the metallicity dispersion of Dra~II casts a doubt of the nature of the satellite, which might well be a globular cluster observed at the very end of its disruption process. Even though challenging, obtaining high signal-to-noise spectra of the faint main-sequence stars of Dra~II currently provides the only hope of unravelling the mystery still surrounding Dra~II.

\section{Acknowledgments}
RI, NL, and NFM gratefully acknowledge funding from CNRS/INSU through the Programme National Galaxies et Cosmologie and through the CNRS grant PICS07708. We gratefully thank the CFHT staff for performing the observations in queue mode. N. F. Martin acknowledges the Kavli Institute for Theoretical Physics in Santa Barbara and the organizers of the ``Cold Dark Matter 2018'' program, during which some of this work was performed. This research was supported in part by the National Science Foundation under Grant No. NSF PHY11-25915. BPML gratefully acknowledges support from FONDECYT postdoctoral fellowship No. 3160510.

Based on observations obtained at the Canada-France-Hawaii Telescope (CFHT) which is operated by the National Research Council of Canada, the Institut National des Sciences de l'Univers of the Centre National de la Recherche Scientifique of France, and the University of Hawaii.

Some of the data presented herein were obtained at the W. M. Keck Observatory, which is operated as a scientific partnership among the California Institute of Technology, the University of California and the National Aeronautics and Space Administration. The Observatory was made possible by the generous financial support of the W. M. Keck Foundation. Furthermore, the authors wish to recognize and acknowledge the very significant cultural role and reverence that the summit of Maunakea has always had within the indigenous Hawaiian community.  We are most fortunate to have the opportunity to conduct observations from this mountain.

The Pan-STARRS1 Surveys (PS1) have been made possible through contributions of the Institute for Astronomy, the University of Hawaii, the Pan-STARRS Project Office, the Max-Planck Society and its participating institutes, the Max Planck Institute for Astronomy, Heidelberg and the Max Planck Institute for Extraterrestrial Physics, Garching, The Johns Hopkins University, Durham University, the University of Edinburgh, Queen's University Belfast, the Harvard-Smithsonian Center for Astrophysics, the Las Cumbres Observatory Global Telescope Network Incorporated, the National Central University of Taiwan, the Space Telescope Science Institute, the National Aeronautics and Space Administration under Grant No. NNX08AR22G issued through the Planetary Science Division of the NASA Science Mission Directorate, the National Science Foundation under Grant No. AST-1238877, the University of Maryland, and Eotvos Lorand University (ELTE).

This work has made use of data from the European Space Agency (ESA)
mission {\it Gaia} (\url{https://www.cosmos.esa.int/gaia}), processed by
the {\it Gaia} Data Processing and Analysis Consortium (DPAC,
\url{https://www.cosmos.esa.int/web/gaia/dpac/consortium}). Funding
for the DPAC has been provided by national institutions, in particular
the institutions participating in the {\it Gaia} Multilateral Agreement.

\newpage

\begin{table*}
\caption{Properties of our spectroscopic sample. Stars A, B and C have a radial velocity of respectively -356.4, -306.2 and -330.5 $\kms$ and are indicated in the column "Member".
\label{tbl-2}}

\setlength{\tabcolsep}{4.5pt}
\renewcommand{\arraystretch}{0.4}
\begin{sideways}
\begin{tabular}{cccccccccccccc}
\hline
RA (deg) & DEC (deg) & $g_0$ & $i_0$ & $CaHK$ & $v_{r} (\kms)$ & $\mu_{\alpha}^{*}$ (mas.yr$^{-1}$) & $\mu_{\delta}$ (mas.yr$^{-1}$) &  S/N & [Fe/H]$_\mathrm{CaHK}$ & $P_{mem}$ & Member & Time\\
\hline

238.34537500 & +64.57397222 & 19.07 $\pm$ 0.01 & 18.62 $\pm$ 0.01 & 19.95 $\pm$ 0.01 & -125.8 $\pm$ 1.3 & $-$4.44 $\pm$ 0.45 & $-$4.02 $\pm$ 0.47 & 23.2 &   ---   & 0.00 &  N & Averaged  \\ \\ 
  & & & & & $-125.0 \pm  1.5$ & & & & & & & 2015 \\ \\
  & & & & & $-129.3 \pm  2.9$ & & & & & & & 2016 \\ \\
238.50220833 & +64.58488889 & 20.05 $\pm$ 0.01 & 19.66 $\pm$ 0.01 & 20.78 $\pm$ 0.01 & -244.0 $\pm$ 3.1 & $-$4.27 $\pm$ 0.86 & 0.74 $\pm$ 0.93 & 9.9 & $-$1.03 $\pm$ 0.15 & 0.37 &  N & 2016  \\ \\ 
238.48954167 & +64.58458333 & 20.59 $\pm$ 0.01 & 19.98 $\pm$ 0.01 & 21.63 $\pm$ 0.03 & -411.9 $\pm$ 5.2 & $-$2.84 $\pm$ 1.44 & $-$5.55 $\pm$ 1.24 & 6.9 & $-$1.07 $\pm$ 0.13 & 0.00 &  N & 2016     \\ \\ 
238.46962500 & +64.61827778 & 20.85 $\pm$ 0.01 & 20.61 $\pm$ 0.01 & 21.28 $\pm$ 0.02 & -328.4 $\pm$ 8.8 & $-$0.04 $\pm$ 3.08 & 2.54 $\pm$ 2.04 & 4.6 & $-$2.50 $\pm$ 0.46 & 0.88 &  Y & 2016    \\ \\ 
238.13162500 & +64.56455556 & 18.85 $\pm$ 0.01 & 18.29 $\pm$ 0.01 & 19.83 $\pm$ 0.01 & -137.2 $\pm$ 1.7 & $-$7.08 $\pm$ 0.32 & 1.52 $\pm$ 0.33 & 25.6 &   ---   & 0.03 &  N & 2016    \\ \\ 
238.32470833 & +64.59466667 & 18.98 $\pm$ 0.01 & 18.89 $\pm$ 0.01 & 19.36 $\pm$ 0.01 & 194.0 $\pm$ 8.1 & 33.07 $\pm$ 0.47 & 42.0 $\pm$ 0.46 & 17.5 &   ---  & 0.00 &  N & 2016    \\ \\ 
238.29208333 & +64.56011111 & 19.36 $\pm$ 0.01 & 18.80 $\pm$ 0.01 & 20.05 $\pm$ 0.01 & -341.6 $\pm$ 0.9 & 1.11 $\pm$ 0.50 & 0.99 $\pm$ 0.49 & 20.5 & $-$2.42 $\pm$ 0.17 & 0.86 &  Y & Averaged  \\ \\ 
  & & & & & $-340.6 \pm  1.1$ & & & & & & & 2015 \\ \\
  & & & & & $-343.7 \pm  1.5$ & & & & & & & 2016 \\ \\
238.22750000 & +64.57172222 & 19.83 $\pm$ 0.01 & 19.38 $\pm$ 0.01 & 20.35 $\pm$ 0.01 & -341.9 $\pm$ 1.7 & 1.64 $\pm$ 0.85 & $-$0.14 $\pm$ 0.87 & 17.3 & $-$2.93 $\pm$ 0.23 & 0.18 &  Y & Averaged  \\ \\ 
  & & & & & $-338.3 \pm  2.8$ & & & & & & & 2015 \\ \\
  & & & & & $-344.1 \pm  2.2$ & & & & & & & 2016 \\ \\
238.17570833 & +64.57013889 & 19.84 $\pm$ 0.01 & 19.34 $\pm$ 0.01 & 20.44 $\pm$ 0.01 & -346.9 $\pm$ 1.6 & 1.45 $\pm$ 0.68 & 0.92 $\pm$ 0.73 & 15.1 & $-$2.57 $\pm$ 0.22 & 0.83 &  Y & Averaged  \\ \\ 
  & & & & & $-345.1 \pm  2.6$ & & & & & & & 2015 \\ \\
  & & & & & $-348.0 \pm  2.0$ & & & & & & & 2016 \\ \\
238.22337500 & +64.55341667 & 20.07 $\pm$ 0.01 & 19.69 $\pm$ 0.01 & 20.53 $\pm$ 0.01 & -341.9 $\pm$ 2.2 & 1.15 $\pm$ 0.84 & 0.46 $\pm$ 0.93 & 13.3 & $-$2.93 $\pm$ 0.23 & 0.93 &  Y & Averaged  \\ \\ 
  & & & & & $-341.0 \pm  2.5$ & & & & & & & 2015 \\ \\
  & & & & & $-345.4 \pm  4.9$ & & & & & & &2016 \\ \\
238.07675000 & +64.59608333 & 20.52 $\pm$ 0.01 & 19.25 $\pm$ 0.01 & 22.50 $\pm$ 0.05 & -95.9 $\pm$ 2.2 & 0.09 $\pm$ 0.73 & $-$7.85 $\pm$ 0.80 & 13.2 &   ---  & 0.00 &  N & Averaged  \\ \\ 
  & & & & & $-94.2 \pm  3.2$ & & & & & & & 2015 \\ \\
  & & & & & $-97.6 \pm  3.1$ & & & & & & &2016 \\ \\
238.21795833 & +64.59575000 & 20.68 $\pm$ 0.01 & 20.45 $\pm$ 0.01 & 21.06 $\pm$ 0.02 & -343.5 $\pm$ 4.6 & 3.81 $\pm$ 1.66 & 1.45 $\pm$ 1.94 & 7.7 & $-$2.77 $\pm$ 0.31 & 0.99 &  Y & Averaged  \\ \\ 
  & & & & & $ -343.6 \pm  14.0$ & & & & & & & 2015 \\ \\
  & & & & & $ -343.5 \pm  4.9$ & & & & & & & 2016 \\ \\
238.19866667 & +64.54908333 & 20.71 $\pm$ 0.01 & 20.41 $\pm$ 0.01 & 21.13 $\pm$ 0.02 & -347.5 $\pm$ 4.8 & 1.91 $\pm$ 1.73 & 0.40 $\pm$ 2.02 & 7.0 & $-$2.72 $\pm$ 0.38 & 0.14 &  Y  & 2016  \\ \\ 
238.15762500 & +64.54386111 & 20.74 $\pm$ 0.01 & 20.49 $\pm$ 0.01 & 21.11 $\pm$ 0.02 & -346.1 $\pm$ 6.9 & $-$0.42 $\pm$ 1.84 & 2.88 $\pm$ 2.03 & 7.2 & $-$2.91 $\pm$ 0.30 & 0.98 &  Y  & 2016   \\ \\ 
238.29766667 & +64.58597222 & 20.90 $\pm$ 0.01 & 20.59 $\pm$ 0.01 & 21.36 $\pm$ 0.02 & -349.7 $\pm$ 6.2 &   ---  &   ---  & 6.6 & $-$2.41 $\pm$ 0.44 & 0.20 &  Y  & 2016   \\ \\ 
238.22362500 & +64.56244444 & 20.94 $\pm$ 0.01 & 20.65 $\pm$ 0.01 & 21.39 $\pm$ 0.02 & -349.7 $\pm$ 5.5 &   ---  &   ---  & 6.6 & $-$2.48 $\pm$ 0.38 & 0.87 &  Y  & 2016   \\ \\ 
238.25691667 & +64.54466667 & 21.05 $\pm$ 0.01 & 20.78 $\pm$ 0.01 & 21.50 $\pm$ 0.02 & -340.0 $\pm$ 8.4 &   ---  &   ---  & 5.4 & $-$2.49 $\pm$ 0.39 & 0.98 &  Y  & 2016   \\ \\ 
238.18845833 & +64.58144444 & 21.62 $\pm$ 0.01 & 21.31 $\pm$ 0.01 & 22.07 $\pm$ 0.03 & -336.3 $\pm$ 12.8 &   ---  &   ---  & 3.6 & $-$2.76 $\pm$ 0.46 & 0.99 &  Y  & 2016   \\ \\ 
238.27908333 & +64.56091667 & 22.71 $\pm$ 0.02 & 21.55 $\pm$ 0.01 & 24.11 $\pm$ 0.17 & -206.3 $\pm$ 5.4 &   ---  &   ---  & 3.3 & $-$2.35 $\pm$ 0.60 & 0.00 &  N  & 2016   \\ \\ 
238.00450000 & +64.55847222 & 18.82 $\pm$ 0.01 & 18.22 $\pm$ 0.01 & 19.91 $\pm$ 0.01 & -47.2 $\pm$ 1.6 & 1.63 $\pm$ 0.30 & $-$2.54 $\pm$ 0.33 & 25.9 &   -   & 0.39 &  N  & 2016   \\ \\ 
238.00004167 & +64.57619444 & 20.37 $\pm$ 0.01 & 18.83 $\pm$ 0.01 & 22.56 $\pm$ 0.05 & -66.7 $\pm$ 2.2 & $-$1.79 $\pm$ 0.53 & $-$10.54 $\pm$ 0.53 & 21.6 &   ---  & 0.00 &  N  & 2016   \\ \\ 
238.05320833 & +64.52075000 & 20.87 $\pm$ 0.01 & 19.56 $\pm$ 0.01 & 22.96 $\pm$ 0.07 & -47.4 $\pm$ 3.4 & $-$6.80 $\pm$ 0.98 & $-$2.43 $\pm$ 0.97 & 12.8 &   ---  & 0.00 &  N  & 2016   \\ \\ 
238.05975000 & +64.55613889 & 22.04 $\pm$ 0.01 & 20.94 $\pm$ 0.01 & 23.80 $\pm$ 0.15 & -253.1 $\pm$ 7.1 &   ---  &   ---  & 5.7 &   ---  & 0.00 &  N  & 2016   \\ \\ 
238.52529167 & +64.59861111 & 16.73 $\pm$ 0.01 & 16.18 $\pm$ 0.01 & 17.74 $\pm$ 0.0 & 3.6 $\pm$ 0.9 & $-$7.16 $\pm$ 0.09 & $-$6.06 $\pm$ 0.09 & 65.4 &   ---  & 0.00 &  N  & 2016  \\ \\ 
238.39033333 & +64.63200000 & 17.60 $\pm$ 0.01 & 16.33 $\pm$ 0.01 & 19.75 $\pm$ 0.01 & -30.9 $\pm$ 1.2 & $-$4.37 $\pm$ 0.10 & $-$2.53 $\pm$ 0.10 & 52.8 &   ---  & 0.00 &  N  & 2016  \\ \\ 
238.34487500 & +64.58811111 & 17.20 $\pm$ 0.01 & 16.60 $\pm$ 0.01 & 18.34 $\pm$ 0.0 & -41.8 $\pm$ 1.2 & $-$6.68 $\pm$ 0.13 & 14.34 $\pm$ 0.14 & 57.4 &   ---  & 0.00 &  N  & 2016  \\ \\ 
238.27975000 & +64.55630556 & 16.93 $\pm$ 0.01 & 16.20 $\pm$ 0.01 & 18.13 $\pm$ 0.0 & -52.0 $\pm$ 1.0 & $-$13.07 $\pm$ 0.19 & $-$0.92 $\pm$ 0.17 & 66.4 & $-$1.12 $\pm$ 0.09 & 0.21 &  N  & 2016  \\ \\ 
238.23100000 & +64.59677778 & 17.76 $\pm$ 0.01 & 17.12 $\pm$ 0.01 & 18.94 $\pm$ 0.0 & -4.1 $\pm$ 0.9 & 0.92 $\pm$ 0.16 & $-$2.94 $\pm$ 0.16 & 47.7 &   ---  & 0.11 &  N  & 2016  \\ \\ 
238.11125000 & +64.57577778 & 17.43 $\pm$ 0.01 & 16.73 $\pm$ 0.01 & 18.73 $\pm$ 0.0 & -68.8 $\pm$ 0.9 & $-$4.60 $\pm$ 0.12 & $-$2.21 $\pm$ 0.13 & 50.4 &   ---   & 0.56 &  N  & 2016  \\ \\ 
238.01183333 & +64.53602778 & 17.96 $\pm$ 0.01 & 17.37 $\pm$ 0.01 & 19.09 $\pm$ 0.01 & -76.7 $\pm$ 1.3 & $-$0.77 $\pm$ 0.18 & $-$3.42 $\pm$ 0.18 & 37.8 &   ---  & 0.00 &  N  & 2016  \\ \\ 
238.56645417 & +64.50524722 & 19.00 $\pm$ 0.01 & 18.42 $\pm$ 0.01 & 20.03 $\pm$ 0.01 & -225.2 $\pm$ 2.3 & $-$2.14 $\pm$ 0.38 & $-$0.71 $\pm$ 0.41 & 5.5 &    ---  & 0.02 &  N & 2015  \\ \\

\end{tabular}
\end{sideways}
\end{table*}

\newpage

\begin{table*}
\caption{Properties of our spectroscopic sample. Stars A, B and C have a radial velocity of respectively -356.4, -306.2 and -330.5 $\kms$ and are indicated in the column "Member". 
\label{tbl-2}}

\setlength{\tabcolsep}{4.5pt}
\renewcommand{\arraystretch}{0.4}
\begin{sideways}
\begin{tabular}{cccccccccccccc}
\hline
RA (deg) & DEC (deg) & $g_0$ & $i_0$ & $CaHK$ & $v_{r} (\kms)$ & $\mu_{\alpha}^{*}$ (mas.yr$^{-1}$) & $\mu_{\delta}$ (mas.yr$^{-1}$) &  S/N & [Fe/H]$_\mathrm{CaHK}$ & $P_{mem}$ & Member & Time\\
\hline

238.51061250 & +64.55532500 & 19.88 $\pm$ 0.01 & 18.28 $\pm$ 0.01 & 22.15 $\pm$ 0.04 & -53.2 $\pm$ 1.5 & 6.87 $\pm$ 0.40 & $-$6.51 $\pm$ 0.42 & 6.3 &   ---  & 0.00 &  N & 2015  \\ \\ 
238.54619167 & +64.52975556 & 20.39 $\pm$ 0.01 & 19.99 $\pm$ 0.01 & 21.11 $\pm$ 0.02 & -165.9 $\pm$ 8.8 & $-$2.78 $\pm$ 1.73 & $-$0.87 $\pm$ 1.18 & 3.0 & $-$1.15 $\pm$ 0.13 & 0.00 &  N & 2015  \\ \\ 
238.55823750 & +64.56613611 & 21.74 $\pm$ 0.01 & 20.00 $\pm$ 0.01 & 23.83 $\pm$ 0.14 & -39.9 $\pm$ 4.7 & $-$3.41 $\pm$ 2.63 & $-$4.42 $\pm$ 2.37 & 5.4 &   ---  & 0.00 &  N & 2015  \\ \\ 
238.41302083 & +64.57990000 & 22.32 $\pm$ 0.01 & 21.37 $\pm$ 0.01 & 23.82 $\pm$ 0.14 & -356.4 $\pm$ 5.5 &   ---  &   ---  & 3.1 & $-$1.34 $\pm$ 0.53 & 0.00 &  N & 2015  \\ \\ 
238.42835417 & +64.57927222 & 22.41 $\pm$ 0.01 & 21.81 $\pm$ 0.02 & 23.06 $\pm$ 0.07 & -306.2 $\pm$ 14.9 &   ---  &   ---  & 3.3 & $-$2.68 $\pm$ 0.63 & 0.00 &  N & 2015  \\ \\ 
238.15138333 & +64.60540000 & 19.44 $\pm$ 0.01 & 18.92 $\pm$ 0.01 & 20.03 $\pm$ 0.01 & -337.6 $\pm$ 2.7 & 0.90 $\pm$ 0.52 & 1.08 $\pm$ 0.54 & 4.1 & $-$2.82 $\pm$ 0.24 & 0.03 &  Y & 2015  \\ \\ 
238.22609167 & +64.59788333 & 19.57 $\pm$ 0.01 & 18.53 $\pm$ 0.01 & 21.26 $\pm$ 0.02 & -84.5 $\pm$ 1.6 & $-$1.76 $\pm$ 0.45 & $-$1.74 $\pm$ 0.46 & 8.6 &   ---  & 0.00 &  N & 2015  \\ \\ 
238.21103333 & +64.57847778 & 20.07 $\pm$ 0.01 & 18.59 $\pm$ 0.01 & 22.24 $\pm$ 0.04 & -3.5 $\pm$ 1.6 & $-$3.22 $\pm$ 0.53 & $-$13.88 $\pm$ 0.55 & 6.7 &   ---   & 0.00 &  N & 2015  \\ \\ 
238.20966667 & +64.61985278 & 20.26 $\pm$ 0.01 & 19.38 $\pm$ 0.01 & 21.58 $\pm$ 0.02 & -182.3 $\pm$ 4.9 & $-$6.00 $\pm$ 0.88 & 3.11 $\pm$ 0.98 & 20.7 & $-$1.62 $\pm$ 0.13 & 0.00 &  N & 2015  \\ \\ 
238.25460417 & +64.54189167 & 20.63 $\pm$ 0.01 & 20.05 $\pm$ 0.01 & 21.66 $\pm$ 0.03 & 9.4 $\pm$ 4.2 & $-$5.38 $\pm$ 1.27 & 2.03 $\pm$ 1.44 & 3.6 & $-$1.02 $\pm$ 0.15 & 0.00 &  N & 2015  \\ \\ 
238.28841250 & +64.58997500 & 21.39 $\pm$ 0.01 & 19.68 $\pm$ 0.01 & 23.71 $\pm$ 0.12 & -61.7 $\pm$ 2.3 & $-$8.05 $\pm$ 1.42 & 1.98 $\pm$ 1.49 & 5.1 &   ---  & 0.00 &  N & 2015  \\ \\ 
238.18912917 & +64.55407222 & 22.21 $\pm$ 0.01 & 21.67 $\pm$ 0.02 & 22.96 $\pm$ 0.07 & -330.5 $\pm$ 5.5 &   ---  &   ---  & 3.6 & $-$2.53 $\pm$ 0.51 & 0.00 &  N & 2015  \\ \\ 
238.03618333 & +64.57617500 & 21.00 $\pm$ 0.01 & 20.73 $\pm$ 0.01 & 21.40 $\pm$ 0.02 & -325.7 $\pm$ 8.9 & 2.71 $\pm$ 2.97 & 0.26 $\pm$ 3.51 & 3.1 & $-$2.90 $\pm$ 0.33 & 0.95 &  Y & 2015  \\ \\ 
238.51431587 & +64.57347835 & 17.98 $\pm$ 0.01 & 17.49 $\pm$ 0.01 & 18.89 $\pm$ 0.0 & -79.9 $\pm$ 1.1 & $-$6.51 $\pm$ 0.19 & 0.49 $\pm$ 0.20 & 13.1 &   ---   & 0.00 &  N & 2015  \\ \\ 
238.49718375 & +64.51780965 & 18.14 $\pm$ 0.01 & 16.31 $\pm$ 0.01 & 20.26 $\pm$ 0.01 & -11.9 $\pm$ 1.1 & $-$68.75 $\pm$ 0.13 & $-$42.7 $\pm$ 0.13 & 14.7 &    ---  & 0.00 &  N & 2015  \\ \\ 
238.43973971 & +64.56761491 & 18.07 $\pm$ 0.01 & 16.91 $\pm$ 0.01 & 20.04 $\pm$ 0.01 & -64.2 $\pm$ 1.3 & 6.83 $\pm$ 0.17 & $-$6.09 $\pm$ 0.17 & 15.1 &   ---  & 0.00 &  N & 2015  \\ \\ 
238.21533400 & +64.60644630 & 15.42 $\pm$ 0.01 & 14.99 $\pm$ 0.01 & 16.29 $\pm$ 0.0 & -54.3 $\pm$ 1.1 & $-$14.42 $\pm$ 0.05 & $-$5.44 $\pm$ 0.05 & 96.1 &   ---  & 0.00 &  N & 2015  \\ \\ 
238.18480597 & +64.61537726 & 17.88 $\pm$ 0.01 & 17.25 $\pm$ 0.01 & 19.09 $\pm$ 0.01 & -35.1 $\pm$ 1.3 & $-$5.66 $\pm$ 0.17 & $-$4.63 $\pm$ 0.17 & 25.8 &   ---   & 0.03 &  N & 2015 \\ \\ 
238.12209840 & +64.60199650 & 18.92 $\pm$ 0.01 & 16.98 $\pm$ 0.01 & 21.15 $\pm$ 0.02 & -0.9 $\pm$ 1.2 & $-$1.69 $\pm$ 0.20 & 11.09 $\pm$ 0.20 & 8.9 &   ---   & 0.00 &  N & 2015  \\ \\ 
238.11487862 & +64.60882637 & 17.89 $\pm$ 0.01 & 16.96 $\pm$ 0.01 & 19.60 $\pm$ 0.01 & -123.9 $\pm$ 1.2 & 4.81 $\pm$ 0.15 & $-$22.05 $\pm$ 0.15 & 11.8 &   ---  & 0.00 &  N & 2015  \\ \\

\end{tabular}
\end{sideways}
\end{table*}

\newcommand{\mnras}{MNRAS}
\newcommand{\pasa}{PASA}
\newcommand{\nat}{Nature}
\newcommand{\araa}{ARAA}
\newcommand{\aj}{AJ}
\newcommand{\apj}{ApJ}
\newcommand{\apjl}{ApJ}
\newcommand{\apjs}{ApJSupp}
\newcommand{\aap}{A\&A}
\newcommand{\aaps}{A\&ASupp}
\newcommand{\pasp}{PASP}


\clearpage

\end{document}